\documentclass{aa}  
\usepackage{graphicx}
\usepackage{natbib}
\usepackage{upgreek}
\usepackage{url}
\usepackage{color}

\bibpunct{(}{)}{;}{a}{}{,} 
\usepackage{txfonts}
\def\degr{\hbox{$^\circ$}}
\def\arcmin{\hbox{$^\prime$}}
\def\arcsec{\hbox{$^{\prime\prime}$}}
\newcommand{\microns}{$\mu$m}
\newcommand{\micron}{$\mu$m}
\newcommand{\pcmm}{cm$^{-2}$}

\newcommand{\Herschel}{\emph{Herschel}}

\newcommand{\mjysr}{MJy\,sr$^{-1}$}
\newcommand{\mjybeam}{mJy\,beam$^{-1}$}
\newcommand{\jybeam}{Jy\,beam$^{-1}$}
\newcommand{\kms}{km\,s$^{-1}$}
\newcommand{\msol}{M$_\odot$}
\newcommand{\lsol}{L$_\odot$}
\newcommand{\mathmsol}{\mathrm{M}_\odot}
\newcommand{\mathlsol}{\mathrm{L}_\odot}
\newcommand{\ccol}{C_\mathrm{col}}

\begin{document} 

   \title{A NIKA view of two star-forming infrared dark clouds:\\
		Dust emissivity variations and mass concentration}

   \author{A.~J.~Rigby\inst{\ref{Cardiff}}
\and  N.~Peretto\inst{\ref{Cardiff}}
\and  R.~Adam \inst{\ref{LPSC},\ref{OCA}}
\and  P.~Ade \inst{\ref{Cardiff}}
\and  P.~Andr\'e \inst{\ref{CEA}}
\and  H.~Aussel \inst{\ref{CEA}}
\and  A.~Beelen \inst{\ref{IAS}}
\and  A.~Beno\^it \inst{\ref{Neel}}
\and  A.~Bracco \inst{\ref{CEA}}
\and  A.~Bideaud \inst{\ref{Neel}}
\and  O.~Bourrion \inst{\ref{LPSC}}
\and  M.~Calvo \inst{\ref{Neel}}
\and  A.~Catalano \inst{\ref{LPSC}}
\and  C.~J.~R.~Clark \inst{\ref{Cardiff}}
\and  B.~Comis \inst{\ref{LPSC}}
\and  M.~De Petris \inst{\ref{Roma}}
\and  F.-X.~D\'esert \inst{\ref{IPAG}}
\and  S.~Doyle \inst{\ref{Cardiff}}
\and  E.~F.~C.~Driessen \inst{\ref{IRAMF}}
\and  J.~Goupy \inst{\ref{Neel}}
\and  C.~Kramer \inst{\ref{IRAME}}
\and  G.~Lagache \inst{\ref{LAM}}
\and  S.~Leclercq \inst{\ref{IRAMF}}
\and  J.-F.~Lestrade \inst{\ref{LERMA}}
\and  J.~F.~Mac\'ias-P\'erez \inst{\ref{LPSC}}
\and  P.~Mauskopf \inst{\ref{Cardiff},\ref{Arizona}}
\and  F.~Mayet \inst{\ref{LPSC}}
\and  A.~Monfardini \inst{\ref{Neel}}
\and  E.~Pascale \inst{\ref{Cardiff}}
\and  L.~Perotto \inst{\ref{LPSC}}
\and  G.~Pisano \inst{\ref{Cardiff}}
\and  N.~Ponthieu \inst{\ref{IPAG}}
\and  V.~Rev\'eret \inst{\ref{CEA}}
\and  A.~Ritacco \inst{\ref{IRAME}}
\and  C.~Romero \inst{\ref{IRAMF}}
\and  H.~Roussel \inst{\ref{IAP}}
\and  F.~Ruppin \inst{\ref{LPSC}}
\and  K.~Schuster \inst{\ref{IRAMF}}
\and  A.~Sievers \inst{\ref{IRAME}}
\and  S.~Triqueneaux \inst{\ref{Neel}}
\and  C.~Tucker \inst{\ref{Cardiff}}
\and  R.~Zylka \inst{\ref{IRAMF}}}
          
   \institute{
School of Physics \& Astronomy, Cardiff University, Queen's Buildings, The Parade, Cardiff, CF24 3AA, UK\\
   \email{RigbyA@cardiff.ac.uk}
	\label{Cardiff}
\and
Laboratoire de Physique Subatomique et de Cosmologie, Universit\'e Grenoble Alpes, CNRS/IN2P3, 53, avenue des Martyrs, Grenoble, France
  \label{LPSC}
\and
  Laboratoire Lagrange, Universit\'e C\^ote d'Azur, Observatoire de la C\^ote d'Azur, CNRS, Blvd de l'Observatoire, CS 34229, 06304 Nice cedex 4, France
  \label{OCA}  
\and
Laboratoire AIM, CEA/IRFU, CNRS/INSU, Universit\'e Paris Diderot, CEA-Saclay, 91191 Gif-Sur-Yvette, France 
  \label{CEA}
\and
Institut d'Astrophysique Spatiale (IAS), CNRS and Universit\'e Paris Sud, Orsay, France
  \label{IAS}
\and
Institut N\'eel, CNRS and Universit\'e Grenoble Alpes, France
  \label{Neel}
\and
Dipartimento di Fisica, Sapienza Universit\`a di Roma, Piazzale Aldo Moro 5, I-00185 Roma, Italy
  \label{Roma}
\and
Institut de Planétologie et d'Astrophysique de Grenoble, Univ. Grenoble Alpes, CNRS, IPAG, 38000 Grenoble, France 
  \label{IPAG}
\and
Institut de RadioAstronomie Millim\'etrique (IRAM), Grenoble, France
  \label{IRAMF}
\and
Institut de RadioAstronomie Millim\'etrique (IRAM), Granada, Spain
  \label{IRAME}
\and
Aix Marseille Universit\'e, CNRS, LAM (Laboratoire d'Astrophysique de Marseille) UMR 7326, 13388, Marseille, France
  \label{LAM}
\and 
LERMA, CNRS, Observatoire de Paris, 61 avenue de l'Observatoire, Paris, France
  \label{LERMA}
\and
School of Earth and Space Exploration and Department of Physics, Arizona State University, Tempe, AZ 85287
  \label{Arizona}
\and 
Institut d'Astrophysique de Paris, CNRS (UMR7095), 98 bis boulevard Arago, F-75014, Paris, France
  \label{IAP}
  }
\authorrunning{A. J. Rigby et al.}
\titlerunning{A NIKA view of two star-forming infrared dark clouds}
   \date{Received ; accepted}
 
 \abstract
	{The thermal emission of dust grains is a powerful tool for probing cold, dense regions of molecular gas in the interstellar medium, and so constraining dust properties is key to obtaining accurate measurements of dust mass and temperature.}
	{By placing constraints on the dust emissivity spectral index, $\beta$, towards two star-forming infrared dark clouds -- SDC18.888-0.476 and SDC24.489-0.689 -- we aim to evaluate the role of mass concentration in the associated star-formation activity.}
	{We exploited the simultaneous 1.2 mm and 2.0 mm imaging capability of the NIKA camera on the IRAM 30 m telescope to construct maps of $\beta$ for both clouds, and by incorporating \textit{Herschel} observations, we created H$_2$ column density maps with 13\arcsec\ angular resolution.}
	{While we find no significant systematic radial variations around the most massive clumps in either cloud on $\gtrsim 0.1$ pc scales, their mean $\beta$ values are significantly different, with $\bar{\beta} = 2.07 \pm 0.09$ (random) $\pm 0.25$ (systematic) for SDC18.888-0.476 and $\bar{\beta} = 1.71 \pm 0.09$ (random) $\pm 0.25$ (systematic) for SDC24.489-0.689. These differences could be a consequence of the very different environments in which both clouds lie, and we suggest that the proximity of SDC18.888-0.476 to the W39 \ion{H}{ii} region may raise $\beta$ on scales of $\sim 1$ pc. We also find that the mass in SDC24.489-0.689 is more centrally concentrated and circularly symmetric than in SDC18.888-0.476, and is consistent with a scenario in which spherical globally-collapsing clouds concentrate a higher fraction of their mass into a single core than elongated clouds that will more easily fragment, distributing their mass into many cores.}
	{We demonstrate that $\beta$ variations towards interstellar clouds can be robustly constrained with high signal-to-noise ratio (S/N) NIKA observations, providing more accurate estimates of their masses. The methods presented here will be applied to the Galactic Star Formation with NIKA2 (GASTON) guaranteed time large programme, extending our analysis to a statistically significant sample of star-forming clouds.}


\maketitle

\section{Introduction}

Despite making up only $\sim$1\% of the mass of the interstellar medium (ISM), dust grains are central to studies of molecular clouds and star formation. Dust grains are a robust tracer of molecular gas density over many orders of magnitude while playing a major role in facilitating the formation of molecules and being an important element of the cooling system of the ISM. Observations of dust extinction or emission -- typically observable from the infrared to millimetre wavelengths -- allow the determination of the masses and temperatures of ISM structures, providing important constraints on the energy balance of objects at various stages of the star-forming process. 

Such mass and temperature determinations depend on the properties of the dust grains, whose composition, size distribution and morphology are usually embodied in the dust emissivity, $\kappa_{\nu}$. In the far-infrared to millimetre regime, where most of the energy from interstellar dust grains is radiated, the emissivity is often described as a single power law:

\begin{equation}
	\kappa_{\nu}=\kappa_0 \left(\frac{\lambda}{\lambda_0}\right)^{-\beta} = \kappa_0 \left(\frac{\nu}{\nu_0}\right)^{\beta},
\end{equation}

\noindent where $\kappa_0$ is the dust opacity at a reference wavelength $\lambda_0$ (or frequency $\nu_0$), and $\beta$ the dust emissivity spectral index. Although it is common practice to assume a uniform dust emissivity law, there is growing evidence that it depends on the environment, through ice coating processes and subsequent dust coagulation, changing both its shape -- $\beta$ -- and normalisation -- $\kappa_0$/$\lambda_0$ \citep[e.g.][]{Ossenkopf+Henning94,Cambresy+01,Ormel+11,Ysard+13}. For example, the presence of large dust particles towards pre-stellar cores was directly highlighted by the observation of `coreshine', that is scattered light at 3--5 \microns\ \citep{Steinacker+10,Pagani+10}, which can only be accounted for by the presence of a population of micron-sized grains. 

The characterisation of $\beta$, and its anti-correlation with dust temperature, has been the subject of a large number of studies \citep[e.g.][]{Dupac+03,Desert+08,Shetty+09a, Smith+12,Sadavoy+13,Juvela+15,Chen+16,Sadavoy+16}. This is most often achieved by fitting the spectral energy distribution (SED) around its emission peak with a modified black-body function whose temperature, dust mass surface density, and emissivity spectral index are left as free parameters. However, moderate noise levels lead to model degeneracy between temperature and spectral index, potentially explaining part of the often-observed anti-correlation between these two parameters \citep{Shetty+09a}. \citet{Chen+16} find that the $T_\mathrm{d}$$-$$\beta$ anti-correlation is significant in Perseus even after accounting for the noise-related degeneracy, with $1.0 \lesssim \beta \lesssim 2.7$. Grain growth in cold dense regions pushes $\beta$ to low values, though the formation of ice mantles can suppress grain growth to increase $\beta$, leading \citet{Chen+16} to propose the sublimation of ice mantles in warm regions as an explanation for the $T_\mathrm{d}$$-$$\beta$ anti-correlation. $\beta$ generally reaches lower values at the locations of protostellar sources. \citet{Dupac+03} and \citet{Yang+Phillips07} found very similar ranges of $\beta$ in nearby star-forming clouds and across a sample of luminous infrared galaxies, respectively. The \citet{Planck11} also reported a $T_\mathrm{d}$$-$$\beta$ anti-correlation that could not be explained by noise alone in Taurus, finding an average value of $\beta \approx 1.8 \pm 0.2$.

Another way of constraining $\beta$ is by observing dust emission at two, or more, wavelengths in the millimetre regime. As one moves closer to the Rayleigh-Jeans tail of the SED, the intensity ratio of these wavelengths becomes less dependent on temperature, greatly reducing the effect of the $T_\mathrm{d}$$-$$\beta$ degeneracy. \citet{Sadavoy+13} used SCUBA-2 maps of the ratio of 450 \micron\ and 850 \micron\ imaging towards several low-mass nearby star-forming regions in the Gould Belt Survey. The authors note that, while the two wavelengths alone can not be used to simultaneously constrain $T_\mathrm{d}$ and $\beta$, relative variations in the latter can be studied if the dust temperature is roughly constant. One issue with these results is that despite the data being at longer wavelengths than most \textit{Herschel} bands, they are not still far enough from the peak of the SED to totally lift the $T_\mathrm{d}$$-$$\beta$ degeneracy, and a difference in dust temperature of 6 K results in a shift in $\beta$ values by $\approx 0.86$. 

More recently, \citet{Bracco+17} used the New IRAM KIDs Array (NIKA) prototype millimetre camera on the Institut de Radioastronomie Millim\'{e}trique (IRAM) 30-m telescope to constrain $\beta$ in the Taurus B213 filament. By using an Abel transform technique in conjunction with the intensity ratio method using 1.2 mm and 2.0 mm continuum images, they were able to simultaneously determine radial temperature and $\beta$ profiles in two low-mass protostellar cores and one pre-stellar core, finding $\beta$ profiles with low central values of $\beta \approx 1.0$ and 1.5 in the protostellar cores, from which $\beta$ increases radially, while the pre-stellar core is consistent with a single $\beta$ value of $2.4$. By observing dust emission simultaneously at 1.2 mm and 2.0 mm, NIKA observations take us closer to the Rayleigh-Jeans regime, allowing a more robust determination of $\beta$. 

Infrared dark clouds (IRDCs) are cold and dense molecular clouds identified in absorption against the mid-infrared background of the Galaxy \citep[e.g.][]{Simon+06,Peretto+Fuller09}. In the past decade or so, IRDCs have been privileged targets for studying the earliest stages of star formation. One key reason for this is that the low number of embedded young stellar objects within IRDCs ensures that the initial conditions for star formation are still imprinted in the IRDC gas properties. By using dendrograms on column density maps (see \citealt{Rosolowsky+08} for an introduction to dendrograms), \citet{Kauffmann+10a,Kauffmann+10b} showed that the mass-size relationship of the identified substructures within a cloud can be linked to the ability of the cloud to form high-mass stars. According to this relation, only a small fraction of known IRDCs have the required conditions to form high-mass stars. These clouds are, therefore, excellent targets to study the early stages of high-mass star formation.

\citet{Peretto+Fuller09} identified a sample of $\sim$11,000 IRDCs in the Galactic plane in the range $10\degr < |\ell| < 65\degr$ and $|b| < 1\degr$ by using 8 \micron\ \textit{Spitzer} GLIMPSE \citep{Benjamin+03} and 24 \micron\ MIPSGAL \citep{Carey+09} data. In this study, we focus on two IRDCs from the \citet{Peretto+Fuller09} catalogue,  SDC18.888-0.476 and SDC24.489-0.689 (hereafter SDC18 and SDC24, respectively), located at $\ell = 18.888\degr $, $b=-0.476\degr$ and $\ell = 24.489\degr$, $b=-0.689\degr$. The two IRDCs also form part of the dense gas survey from Peretto et al. (in prep.) and were targeted in this study because of their differences in morphology, masses, sizes, and mid-IR environment. Systemic velocities in N$_2$H$^+$ (1$-$0) were measured by Peretto et al. (in prep) to be +66.3 \kms\ and +48.1 \kms, yielding kinematic distances  of $4.38 \pm 0.27$ kpc and $3.28 \pm 0.34$ kpc, respectively, adopting the \citet{Reid+09} Galactic rotation model. SDC18 lies on the edge of the W39 \ion{H}{II} region, a region of widespread active high-mass star formation \citep[HMSF;][]{Kerton+13}, and harbours an Extended Green Object at its centre \citep{Cyganowski+08}. By contrast, SDC24 is relatively secluded, with no \ion{H}{II} region identified in the area. Only an outflow, identified by H$_{2}$ knots, has been associated with a young stellar object (YSO) located towards the centre of the cloud \citep{Ionnidis+Froebrich12}.

Here, we use NIKA data to constrain, for the first time, $\beta$ variations within two infrared dark clouds (IRDCs) and to use these results to characterise the mass concentration within the IRDCs as a function of their global properties. This paper is divided into the following Sections. In Section \ref{sec:Observations} we describe the observations, data reduction and calibration. We describe our methodology for the $\beta$ determination in Section \ref{sec:Betamaps} and our mass determination is presented in Section \ref{sec:Mass}. A discussion of the implications of the results follows in Section \ref{sec:Discussion}, and we summarise our results to conclude in Section \ref{sec:Summary}.

\section{Observations} \label{sec:Observations}

\subsection{NIKA data} \label{sec:NIKA}

\begin{figure*}
	\centering
	\includegraphics[width=17cm]{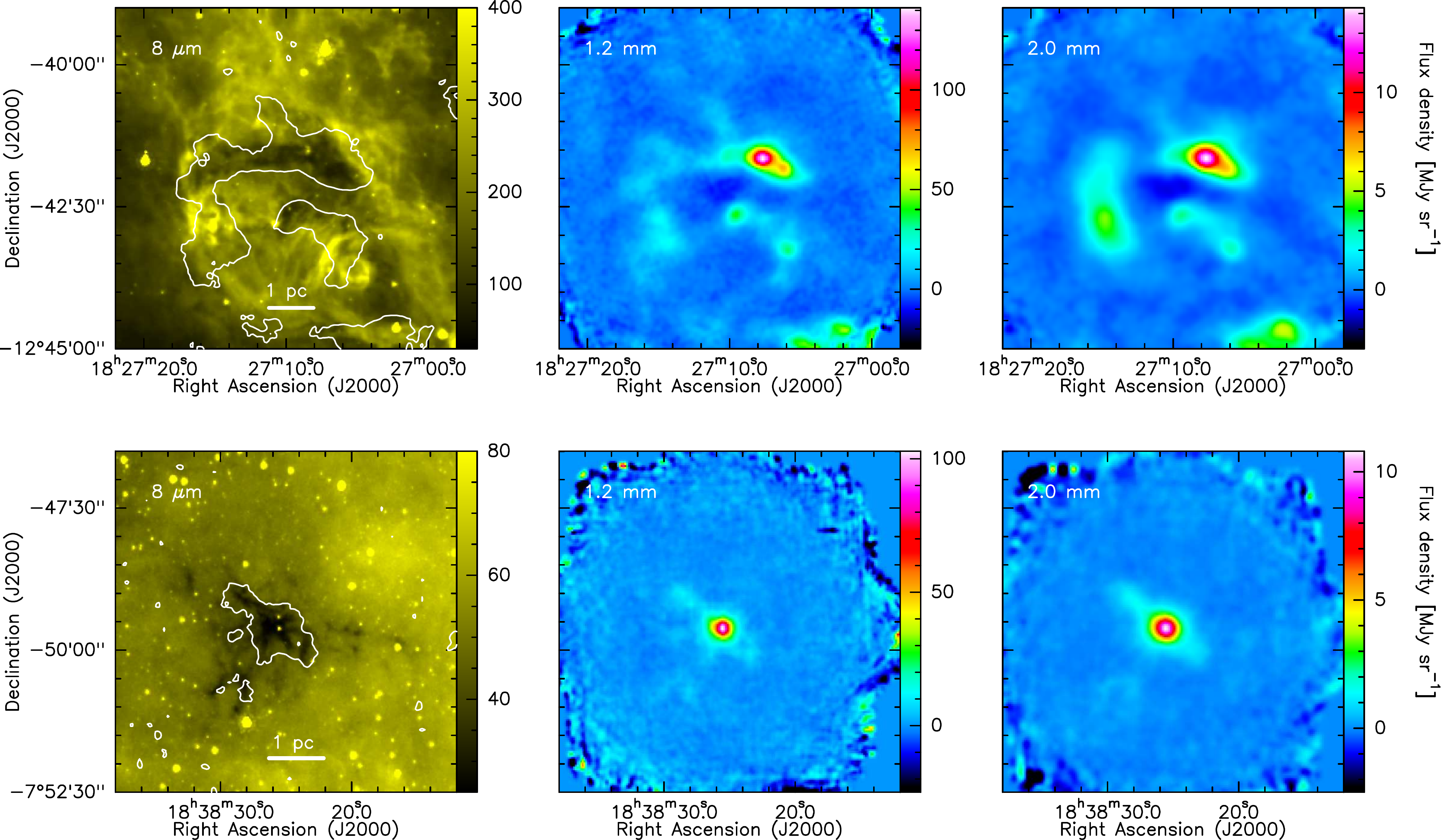} 
    \caption{Top row: images of SDC18, with \textit{Spitzer} 8 \micron\ images in yellow, followed by the 1.2 mm and 2.0 mm NIKA images, slightly smoothed to 13 and 20 arcseconds resolution, respectively. Bottom row: as above for SDC24. The contours on the 8 \micron\ images show the where S/N=3 from the 13-arcsecond resolution 1.2 mm images.}
	\label{fig:rawdata}
\end{figure*}

We used NIKA \citep{Monfardini+10,Monfardini+11,Catalano+14} on the IRAM 30 m telescope to map SDC18 and SDC24 simultaneously in the 1.2 mm and 2.0 mm continuum wavebands. The observations were made on November 16th 2014 as part of the NIKA Open Pool 2 campaign and reduced following the procedures described in \citet{Catalano+14} and \citet{Adam+14}. Mean zenith opacities at 225 GHz for SDC18 and SDC24 were $\tau_{225} =0.23$ and 0.20, and the mean elevations were 45\degr\ and 35\degr, respectively. 

The NIKA maps were calibrated using observations of Uranus at reference frequencies of 260 GHz and 150 GHz for the 1.2 mm and 2.0 mm bands, respectively. Since the IRDCs have SEDs which differ to that of Uranus, a colour correction must be applied that accounts for the response of the two NIKA bandpasses as a function of frequency, effectively converting the fluxes to monochromatic values at the nominal frequencies. The model of Moreno (2010)\footnote{ftp://ftp.sciops.esa.int/pub/hsc-calibration/PlanetaryModels/ESA2/} was adopted for the SED of Uranus. The colour correction assumes a source SED of the form $f(\nu/\nu_0) = (\nu/\nu_0)^\alpha$, and we adopted a value of $\alpha= 3.8$ for sources described by a modified black-body model in the Rayleigh-Jeans limit, with dust emissivity spectral index $\beta = 1.8$, which is the average value in the Milky Way disk \citep{Planck11}. The colour correction factors were determined to be $\ccol = 0.971$ and 0.917 for the 1.2 mm and 2.0 mm bands, where $S_\mathrm{corr} = S_\mathrm{obs} \times \ccol$ based on this assumed spectral shape.

Uranus observations were also used to determine the profile of the primary telescope beam. The main beam of the telescope is well fitted by a Gaussian distribution with a full-width half-maximum (FWHM) of 12.0\arcsec\ and 18.2\arcsec\ for the 1.2 mm and 2.0 mm wavebands, respectively \citep{Ruppin+17}, though we note that for extended emission, there are significant non-Gaussian sidelobes at large radii which should be accounted for. The Uranus observations were also used to determine the effective solid angle $\Omega_\mathrm{A}$ of the 1.2 mm and 2.0 mm beams, which were measured to be $4.51 \times 10^{-9}$ sr and $10.99 \times 10^{-9}$ sr, respectively, allowing the raw maps to be converted from their intrinsic units of \jybeam\ to \mjysr. We summarise the basic properties of the NIKA images, and those from the other observations used in this study in Table \ref{tab:observations}. The NIKA images are presented in Fig. \ref{fig:rawdata} alongside 8 \micron\ images from the \textit{Spitzer} survey GLIMPSE in which the IRDC components were originally identified by \citet{Peretto+Fuller09}.

\begin{table*}
	\caption{Details of the wavebands of the various single-dish observations used in this study. For the LABOCA and NIKA images, we detail the effective beam solid angles used to convert the images into units of MJy sr$^{-1}$, though we note that the Hi-GAL images are supplied in units of MJy sr$^{-1}$. The nominal wavelengths or frequencies recovered after colour correction are denoted in each case are denoted by an asterisk, and bandwidths are quoted in corresponding units.}
	\label{tab:observations}
	\centering
	\begin{tabular}{l l l l l l l l}
	\hline\hline 
	Facility			& Instrument	&	$\lambda_0$	& $\nu_0$	& FWHM		& $\Omega_\mathrm{A}$ 	& Bandwidth			& $C_\mathrm{col}$	\\ 
						& 				& 	(\micron)   & (GHz)		& (arcsec)	& ($10^{-9}$ sr)	& (\micron, GHz)	&  					\\
	\hline \\[-2ex] 
	\textit{Herschel} 	& PACS			&	70* 		& 4280		& 6.0$^a$	& --					& 60--85$^c$	 	& 1.00				\\ 
						& PACS			&	160* 		& 1870		& 12.0$^a$	& --					& 130--210$^c$	 	& 1.01	 			\\ 
						& SPIRE			&	250*		& 1200		& 18.0		& --					& 211--290$^c$		& 1.02				\\ 
            			& SPIRE			&	350*		& 857		& 24.0		& --					& 297--405$^c$	 	& 1.01				\\ 
    					& SPIRE			&	500*		& 600		& 35.0		& --					& 409-611$^c$	 	& 1.03				\\ 
    \\[-2ex] \hline \\[-2ex]
	APEX				& LABOCA		&	869			& 345*		& 19.2		& 12.17$^b$				& 313--372$^c$		& 0.988				\\
    \\[-2ex] \hline \\[-2ex]
	IRAM				& NIKA 			&	1153		& 260*		& 12.0		& 4.51					& 196--273$^d$		& 0.971				\\ 
	30-m				& NIKA			&	1999		& 150*		& 18.2		& 10.99					& 127--171$^d$		& 0.917				\\ 
	\hline 
	\end{tabular}
    \tablefoot{\\
	\tablefoottext{a}{We note that the 70 \micron\ and 160 \micron\ Hi-GAL images do not achieve the nominal beam size, but instead have PSFs that are elongated in the scan direction, and are measured to be $5.8\arcsec \times 12.1\arcsec$ and $11.4\arcsec \times 13.4\arcsec$, respectively. See \citet{Molinari+16} for more details.}\\
    \tablefoottext{b}{We adopt the LABOCA total beam area for extended sources. See http://www.apex-telescope.org/bolometer/laboca/calibration/.}\\
	\tablefoottext{c}{The PACS, SPIRE and LABOCA waveband edges are defined as half of the average in-band transmission \citep{Siringo+09,Poglitsch+10,Swinyard+10}.}\\
	\tablefoottext{d}{The NIKA waveband edges contain ninety percent of the total transmission \citep{Adam+14}.}
    }
\end{table*}

The rms noise levels in the NIKA images vary according to position as a consequence of the scanning pattern and the reduced integration time per pixel at the image edges. However, noise levels are relatively constant within $\sim 1.8\arcmin$ of the centre in SDC18, or within $\sim 1.2\arcmin$ of the centre of SDC24, which is approximately where the main emission features lie. The central rms noise values in both pairs of 1.2 mm and 2.0 mm images for SDC18, once slightly smoothed to effective angular resolutions of 13\arcsec\ and 20\arcsec\ (and before application of the colour corrections) are 7.1 \mjybeam\ and 2.2 \mjybeam, respectively, while for SDC24 the central rms noise values are 4.8 \mjybeam\ and 0.9 \mjybeam, respectively. Details of the rms noise levels and mean background levels in units of \mjysr\ are presented in Table \ref{tab:noise}.

\begin{table}
	\caption{Statistics of NIKA images under various levels of smoothing. Individual images are denoted by the approximate central wavelength $\lambda$ in the corresponding waveband, with an effective resolution element of FWHM $\theta$ after smoothing. Smoothing using kernels denoted `G' are Gaussian, while those denoted `P' used the PSF-matching kernel described in Sect \ref{sec:crkernel}, and where no kernel is listed, measurements of the raw (unsmoothed) data are presented. We also list the mean integration time, $t_\mathrm{obs}$ per 2\arcsec\ pixel in the central 1.8\arcmin\ in SDC18 and 1.2\arcmin\ in SDC24 used to calculate the effective rms noise level $\sigma$ and mean background level $\mu$.}
	\label{tab:noise}
	\centering
	\begin{tabular}{c c c c c r}
	\hline\hline 
	Source	& $\lambda$	& $\theta$ 	& Kernel & $t_\mathrm{obs}$ & $\mu \pm \sigma \ \ \ \ $	\\ 
			&  (mm)		& (\arcsec) & 		 & (s/pix)			& (\mjysr) \\
	\hline \\[-2ex] 
   	SDC18	& 1.2 	& 12.0	& --	& 20.6	& $-0.10 \pm 2.35$	\\
			& 1.2 	& 13.0	& G		& 20.6	& $0.14 \pm 1.57$	\\
		 	& 1.2 	& 20.0	& P		& 20.5	& $0.17 \pm 1.15$	\\
	 		& 1.2 	& 20.0	& G		& 20.5	& $0.17 \pm 1.16$	\\
			& 2.0 	& 18.2	& -- 	& 15.9	& $0.01 \pm 0.29$	\\
			& 2.0 	& 20.0	& G 	& 15.9	& $0.02 \pm 0.20$	\\
	\\[-2ex]	
	\hline
	\\[-2ex]
   	SDC24 	& 1.2 	& 12.0	& --	& 17.9	& $-0.03 \pm 2.13$	\\
		 	& 1.2 	& 13.0	& G		& 17.8	& $0.15 \pm 1.01$	\\
		 	& 1.2 	& 20.0	& P		& 17.8	& $0.12 \pm 0.62$	\\
			& 1.2 	& 20.0	& G		& 17.8	& $0.12 \pm 0.60$	\\
			& 2.0 	& 18.2	& --	& 13.6	& $0.00 \pm 0.28$	\\
		 	& 2.0 	& 20.0	& G		& 13.6	& $0.00 \pm 0.09$	\\

	\hline 
	\end{tabular}
    
\end{table}

Extended emission on scales larger than the $\sim 2$\arcmin\ field of view is removed during the atmospheric decorrelation of the raw NIKA timelines, in which it is impossible to distinguish between astrophysical and atmospheric signal. Consequently, the 1.2 mm and 2.0 mm NIKA images of SDC18, which lies in a region of bright extended emission (as seen in the \textit{Spitzer} images of Fig. \ref{fig:rawdata}) contain a region of negative emission towards the centre. The spatial filtering has the effect of reducing the flux in every pixel, and while the effect is small in regions where the emission is compact, such as towards the emission peaks, it can be large in diffuse regions. We discuss the impact of this filtering upon our results in the relevant Sections of this paper.

\subsection{Herschel data}

Data from the \Herschel\ infrared Galactic Plane Survey \citep[Hi-GAL;][]{Molinari+10,Molinari+16}, consisting of PACS \citep{Poglitsch+10} imaging at 70 and 160 \micron, and SPIRE \citep{Griffin+10} imaging at 250, 350, and 500 \microns\ were also used. We used the DR1 versions of the photometric maps, described in \citet{Molinari+16}, which have been reduced using the ROMAGAL data reduction pipeline, described in \citet{Traficante+11}. As part of DR1, the reduced maps have been calibrated to a common zero level with offsets determined by comparison to all-sky maps from \textit{Planck} and IRIS, following the process outlined in \citet{Bernard+10}.

The \Herschel\ images are calibrated assuming a flat source spectrum, that is $I_\nu \propto \nu^{-1}$. We adopted the colour corrections determined by \citet{Sadavoy+13}, who found $\ccol = 1.01, 1.02, 1.01$ and 1.03 for the 160, 250, 350 and 500 \micron\ bands, respectively, for extended sources with dust temperatures of $T_\mathrm{d} \approx 10-15$ K and dust emissivity spectral indices of $\beta \approx 1.5-2.5$. It is not necessary to apply any colour correction to the 70 \micron, given its minor role in this study. We adopt absolute calibration uncertainties at the level of 5\% for PACS and 4\% for SPIRE imaging \citep{Molinari+16} throughout this study.

\subsection{ATLASGAL data}

In addition to the NIKA and Hi-GAL continuum imaging, we made use of the ATLASGAL survey \citep{Schuller+09} whose observations were made using the LABOCA camera \citep{Siringo+09}. These data have an angular resolution of 19.2\arcsec, and are calibrated on planet observations with a spectrum of $I_\nu \propto \nu^{2}$ and a reference frequency of 345 GHz. We applied a colour correction factor of 0.988 to the data, following \citet{Csengeri+16}, in order to match the SED assumption made to calculate the NIKA colour correction in Sect. \ref{sec:NIKA} -- a modified black-body with $\beta = 1.8$. All colour corrections applied to the various data in this study are summarised in Table \ref{tab:observations}.

\subsection{MAGPIS data}

Data from the Multi-Array Galactic Plane Imaging Survey \citep[MAGPIS;][]{Helfand+06} were also used in this work. The 20 cm continuum MAGPIS data were taken from the Very Large Array (VLA) in pseudocontinuum mode in B, C and D-configurations, providing a sensitivity of 1--2 mJy with a 6\arcsec-resolution synthesised beam. The Image Cutout service\footnote{https://third.ucllnl.org/cgi-bin/gpscutout} was used to extract $10\arcmin \times 10\arcmin$ 20cm continuum images covering the SDC18 and SDC24 NIKA fields, which have rms noise values of 0.2 mJy beam$^{-1}$ and 0.3 mJy beam$^{-1}$, respectively.

\section{Constructing $\beta$ maps using NIKA data} \label{sec:Betamaps}

NIKA observations provide a powerful complement to \Herschel\ data by providing greater angular resolution at wavelengths longer than 500 \micron\ in the spectral energy distributions of thermal dust emission. These can be used to better constrain the spectral index of the dust emissivity, thus providing tighter constraints on the dust masses, as well as improving the identification of cold compact sources and underlying dust structure in each IRDC. In this Section we describe the steps necessary to construct maps of $\beta$ using NIKA data and discuss the nature of the related uncertainties.

\subsection{Clump properties} \label{sec:SED}

Both SDC18 and SDC24 contain one dominant compact clump which contains the majority of the dust emission, centred on $\ell$ = 18.887\degr, b = -0.474\degr and $\ell$ = 24.488\degr, b = -0.691\degr, respectively. To acquire an initial estimate of the properties of these clumps, their spectral energy distributions (SEDs) were measured, and fitted to a modified black body model, which has a flux density $I_\nu$ at a frequency $\nu$ given by:

\begin{equation} \label{eq:greybody}
I_{\nu}=N_{\mathrm{H}_2} \mu_{\mathrm{H}_2} m_\mathrm{H} \kappa_0 \left(\frac{\nu}{\nu_0}\right)^{\beta} B_{\nu} (T_\mathrm{d}),
\end{equation}

\noindent where $N_{\mathrm{H}_2} $ is the H$_2$ column density, $\mu_{\mathrm{H}_2}$ is the mean molecular weight per hydrogen molecule which has a value of 2.8 for molecular gas with a relative Helium abundance of 25\%, $m_\mathrm{H}$ is the mass of a hydrogen atom, $ \kappa_0$ is the specific dust opacity at the reference frequency $\nu_0$, $\beta$ is the dust emissivity spectral index, and $B_{\nu} (T_\mathrm{d}) $ is the Planck function evaluated at the dust temperature $T_\mathrm{d}$. It is important to bear in mind that these quantities are necessarily line-of-sight averages. We stress that the SED-fitting procedure provides an initial estimate of the clump properties, and in particular the dust temperature which will be required to produce $\beta$ maps in Section \ref{sec:betamap}. It is, therefore, subject to the $T_\mathrm{d}$-$\beta$ degeneracy though this should be expected to have the same systematic effect for both sources.

To supplement the 1.2 mm and 2.0 mm NIKA images, Hi-GAL 160 \micron\ and 250 \micron\ images as well as ATLASGAL 870 \micron\ images, which all have comparable or better angular resolution than the NIKA imaging, were used to constrain the SEDs. Before constructing SEDs we resampled the Hi-GAL and ATLASGAL images on to the same 2\arcsec-pixel grid as the NIKA data. As a consequence all images were largely oversampled, but this was necessary to check for astrometric consistency between the different telescopes. A systematic shift of +4\arcsec\ in declination was found in the Hi-GAL data compared to the NIKA and ATLASGAL data, which was estimated by comparing the emission centroid of the three main emission peaks in the Herschel 250 \micron\ and NIKA 1.2 mm images, and the Hi-GAL images were accordingly shifted by $\Delta \delta=-4$\arcsec\ to account for this discrepancy.

As space-based observations, the Hi-GAL images recover intensity on all angular scales, in contrast to the ground-based ATLASGAL and NIKA imaging which necessarily lose flux from large-scale emission as a result of the filtering of atmospheric fluctations in the data reduction processes. Comparison of the fluxes from the ground- and space-based images should ideally contain the same sensitivity to all scales of emission, and so we constructed a transfer function for NIKA to apply equivalent spatial filtering to the \textit{Herschel} observations. To achieve this, a series of synthetic images with defined angular power spectra were generated and processed through the same reduction pipeline as was used for the raw observations of SDC18 and SDC24. The power spectra of the output maps, computed with POKER \citep{Ponthieu+11}, were then compared to those of the input maps in order to construct an average transfer function for both NIKA wavelengths, which we display in Fig. \ref{fig:transferfunction}. After application of the appropriate transfer function, the Herschel images can be regarded as being equivalently spatially filtered. We applied the 1.2 mm and 2.0 mm NIKA transfer functions to the 160 \micron\ and 250 \micron\ Herschel images, respectively, as these wavebands have well-matched angular resolutions. We applied no further filtering to the ATLASGAL data, noting that LABOCA will have recovered more extended emission since it has a larger field of view than NIKA ({with a diameter of $11\arcmin$ compared to $2\arcmin$, respectively), and therefore will contain an additional flux contribution from extended emission. 

\begin{figure} 
	\centering
    \includegraphics[width=\columnwidth]{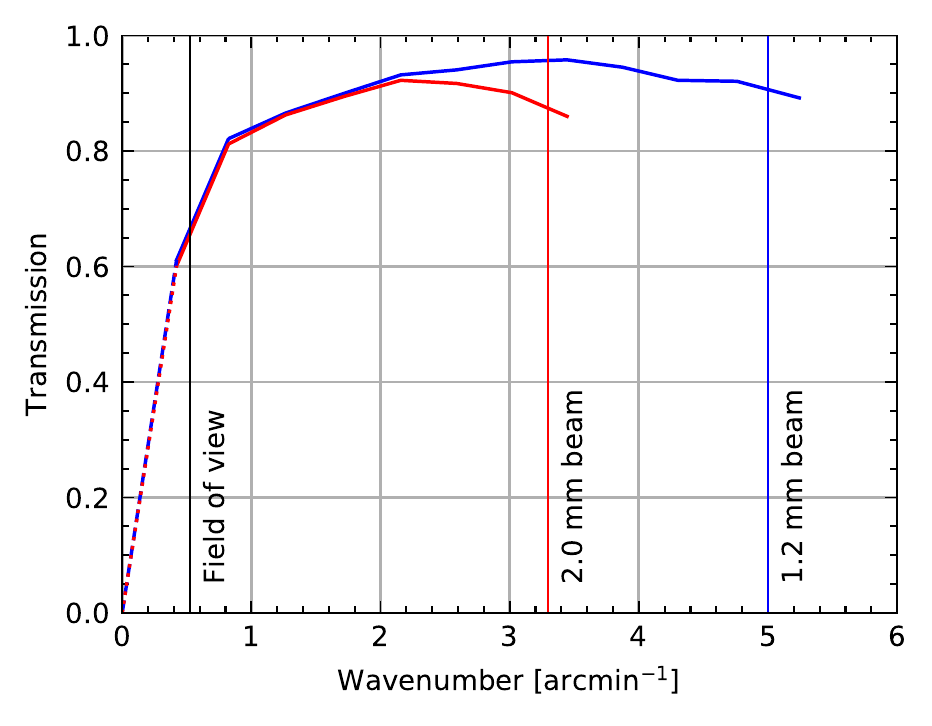}
    \caption{Transfer function for NIKA data reduced in this study. The vertical lines identify spatial frequencies associated with the 1.9\arcmin\ NIKA 2.0 mm field of view (for clarity we do not also show the 1.8\arcmin\ field of view at 1.2 mm), as well as the corresponding FWHM beam sizes in both wavebands. The dashed part of the transfer functions have been extrapolated to a zero-level of transmission at a wavenumber of 0 arcmin$^{-1}$.}
	\label{fig:transferfunction}
\end{figure}

\begin{figure}
	\centering
	\includegraphics[width=\hsize]{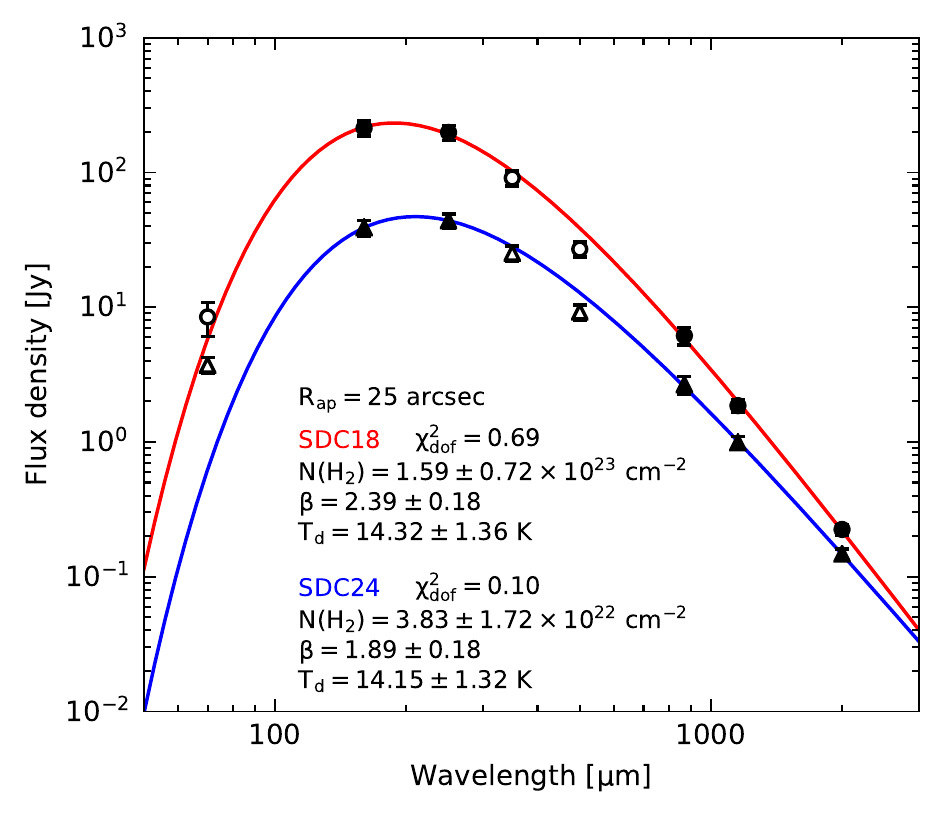}
	\caption{SED fits for the main clumps in SDC18 (red) and SDC24 (blue), along with the best fit parameters. The aperture photometry measurements are shown as filled circles for SDC18 and filled triangles for SDC24. We show the measurements at 70 \micron, 350 \micron\ and 500 \micron\ that were not used in the fitting procedure as empty symbols.}
	\label{fig:SED}
\end{figure}

The Python package \textit{photutils} was used to carry out aperture photometry on each clump, using an aperture with a 25\arcsec\ radius, equal to $\sim 3$ times the standard deviation of the LABOCA beam, which has the lowest angular resolution. An annular sky aperture was used with inner and outer radii equal to 1.5 and 2.0 times that of the source aperture, respectively. The rms noise level was determined in each image after applying 50 iterations of sigma-clipping with a 2.5 sigma cut-off. At each wavelength, the rms uncertainties only contribute a relatively small uncertainty to the integrated fluxes, whose errors are dominated by the calibration uncertainties. The calibration uncertainties were taken to be 5\% for the unfiltered \Herschel\ PACS band at 160 \micron\ \citep{Molinari+16}, 4\% for the unfiltered \Herschel\ SPIRE 250 \micron\ band \citep{Molinari+16}, 15\% for the 870 \micron\ ATLASGAL image \citep{Schuller+09}, and 11\% and 9\% for the 1.2 mm and 2.0 mm NIKA bands, respectively \citep{Ruppin+17}. We applied no aperture corrections to the aperture photometry measurements on the basis that the sources are extended.

After our application of the NIKA transfer function to the Hi-GAL data, we must increase the flux uncertainties measured in the apertures accordingly. The absolute calibration uncertainties for the NIKA images, determined by \citet{Catalano+14}, contains a 5\% contribution from the spatial filtering in the case of point sources. Since we are looking at dense cores that are marginally resolved in the NIKA data, the spatial filtering must contribute a minimum of 5\% uncertainty, but should be of a similar order of magnitude. By comparing the mean aperture flux in the filtered and unfiltered 160 \micron\ and 250 \micron\ \textit{Herschel} bands, weighted by the 1.2 mm intensity per pixel, we conservatively estimate that the NIKA transfer function contributes an additional fractional uncertainty of 12\% to the filtered \textit{Herschel} images.

The SED fitting was performed using the Python \textit{SciPy} routine \textit{curve\_fit}, which performs a $\chi^2$ minimisation via the Levenberg--Marquardt algorithm. Figure \ref{fig:SED} shows the measured SEDs for SDC18 and SDC24 along with the modified black body fits along with the best fit parameters. The dust temperatures are consistent with each other, with a value of $14.32 \pm 1.36$ K in SDC18 and $14.15 \pm 1.32$ in SDC24. The average column densities per 2\arcsec\ pixel within the 25\arcsec\ apertures are $1.6 \pm 0.7 \times 10^{23}$ \pcmm\ and $3.8 \pm 1.7 \times 10^{22}$ \pcmm\ for SDC18 and SDC24, respectively, which yield masses of $3200 \pm 1500$ \msol\ and $430 \pm 210$ \msol. At distances of 4380 pc and 3280 pc, a 25\arcsec\ aperture corresponds to $\sim$0.5 pc and $\sim$0.4 pc, respectively. In Fig. \ref{fig:SED} we also include the aperture photometry measurements in the 70 \micron\ band, though it should be noted that these points were not used in the SED-fitting as 70 \micron\ emission often contains non-equilibrium dust heating from embedded young stellar objects \citep{Dunham+08}. We were unable to spatially filter the 350 \micron\ and 500 \micron\ \textit{Herschel} images in a way that is consistent with the NIKA data points since their beamsizes are considerably larger (indeed, the close match of the resolutions of the 160 \micron\ and 250 \micron\ \textit{Herschel} images with the 1.2 and 2.0 mm NIKA images is fortuitous). We include the photometric measurements from these images in Fig. \ref{fig:SED} after applying the 2.0 mm NIKA transfer function for clarity, but note that the filtering is necessarily too harsh due to its determination from a smaller beamsize, which results in underestimated fluxes. We note that the $\chi^2$ per degree of freedom value for SDC24 is rather low (see Fig. \ref{fig:SED}), and we can only recover a value of unity by artificially decreasing the uncertainties resulting from the absolute calibration to an unrealistic level. We interpret this as an indication that there may be some level of correlation between uncertainties that we have not accounted for in this approach, and the two NIKA wavebands must be the prime candidate. However, investigating potential correlations between the NIKA calibrations is beyond the scope of this paper.

\subsection{Constructing a PSF-matching convolution kernel} \label{sec:crkernel}

The 1.2 mm and 2.0 mm NIKA wavebands have different angular resolutions and so a common resolution must be achieved before comparison of the intensities of pixels covering the same areas of sky can be used to derive physical quantities. A standard method of achieving this is to use Gaussian smoothing kernels, where the quadrature sum of the FWHMs of the beam and the smoothing kernel is equal to that of the desired common resolution. In principle, this method will work well for telescope beams that follow a Gaussian profile, but the two NIKA beams show a significant departure from a Gaussian profile in the sidelobes. The NIKA sidelobes can contain as much as 43\% of the power \citep{Catalano+14} in the point spread function (PSF), and neglect of this matter may introduce artificial features in subsequent analyses.

To effectively match the PSFs in maps made in the two NIKA wavebands, we created a PSF-matching kernel that, when applied to the 1.2 mm imaging, will result in an image with the same effective PSF as the 2.0 mm image. The {\it psf.matching} subpackage of the {\it photutils} \citep{Bradley+16} Python package was used to generate a convolution kernel using the ratio of the fast Fourier transforms (FFTs) of the 1.2 mm and 2.0 mm PSFs \citep[e.g.][]{Gordon+08, Aniano+11,Pattle+15}. The NIKA PSFs were constructed by taking the mean of three beam maps made from observations of Uranus on the nights of the 19th and 20th of February 2014 during the preceding observing campaign. Prior to taking the mean value of each pixel in the beam maps, the maps were normalised to the peak intensity to ensure a consistent scaling and then circularly averaged following the method described in Section 4.4 of \citet{Aniano+11}. The PSFs have been spatially filtered to reduce the impact of high frequency noise arising from finite numerical precision. We used a Top Hat window function to exclude spatial frequencies with an intensity of less than 0.5\% of the maximum value  of the 1.2 mm PSF's FFT. In this way, spatial frequencies higher than 40\% of the highest frequency (corresponding to a 2 pixel-wavelength oscillation) were removed.

\begin{figure}
	\centering
	\includegraphics[width=\columnwidth]{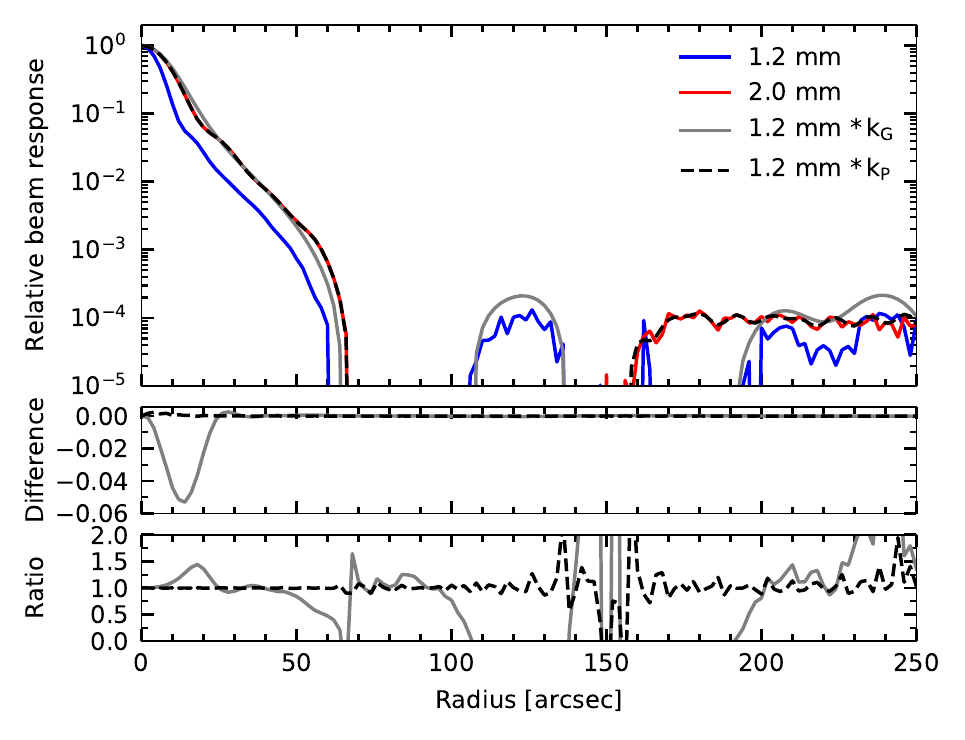}
	\caption{Profiles of the 2.0 mm NIKA beam (red line), alongside the 1.2 mm NIKA beam  before (blue line) and after convolution with the Gaussian (grey line) and PSF-matching kernels (black dashed line). The bottom two panels show the differences between, and the ratios of, the two 1.2 mm beam profiles smoothed to the 2.0 mm resolution using the Gaussian and PSF-matching kernels and the 2.0 mm profile. Buckling of the panels of the primary mirror show up as features at $\sim 2$\arcmin\ and 3.5\arcmin\ in the average 1.2 mm and 2.0 mm beam profiles, respectively.}
	\label{fig:kernels}
\end{figure}

Radial profiles of the circularly-averaged NIKA beams are presented in Fig. \ref{fig:kernels}, along with profiles of 1.2 mm beam after convolution with the PSF-matching kernel and a Gaussian kernel for comparison. The middle and lower panels show the difference and ratio between the 1.2 mm profiles convolved using the two methods and the 2.0 mm profile. The middle panel shows that the PSF-matching kernel is able to map the 1.2 mm PSF on to the 2.0 mm PSF extremely well, and has a better than 1\% agreement, assuming circular symmetry, up to a radius of 4\arcmin. By comparison, resolution-matching using a purely Gaussian kernel results in a significant departure at radii up to $\sim 20$\arcsec\ that could be particularly problematic for compact sources. Indeed, in the lower panel, showing the ratio between the convolved 1.2 mm profiles and the 2.0 mm profile, it can be seen that the Gaussian convolution method causes an artificial change in the ratio of 50\% at around 20\arcsec, though while there are much larger deviations at larger radii, the power in the beam beyond 60\arcsec\ is so low as to have no effect on these images.

Finally, NIKA maps of SDC18 and SDC24 were produced at a common resolution of 20 arcsec for both 1.2 mm and 2.0 mm wavebands. To achieve this, the 1.2 mm maps were first convolved with the PSF-matching kernel to the angular resolution of the 2.0 mm images. A Gaussian fit to the 2.0 mm PSF yields a FWHM of 18.2 arcsec, and so a Gaussian smoothing with an 8.3 arcsec-FWHM kernel was applied to both the convolved 1.2 mm and raw 2.0 mm images at this point to achieve a common resolution of 20 arcsec. Smoothing to 20\arcsec\ is necessary because, while the 1.2 mm image is already very smooth at this point, having been convolved to $\sim 18.2\arcsec$ already, the 2.0 mm image contains significant noise in the background at its native resolution. In addition, the ancillary imaging from Hi-GAL, ATLASGAL and MAGPIS may also be used after smoothing to 20 arcseconds.

\subsection{Intensity ratio maps} \label{sec:ratiomap}

Following Eq. \ref{eq:greybody}, the ratio of intensities at 1.2 mm and 2.0 mm, $I_1$ and $I_2$, can be expressed as:

\begin{equation} \label{eq:ratio}
	\frac{I_1}{I_2} = \left( \frac{\nu_1}{\nu_2}\right)^\beta \frac{B_1 (T_\mathrm{d})}{B_2 (T_\mathrm{d})}.
\end{equation}

\noindent where $B_1(T_\mathrm{d})$ and $B_2(T_\mathrm{d})$ are Planck functions evaluated at 1.2 mm and 2.0 mm, respectively, corresponding to reference frequencies $\nu_1$ and $\nu_2$. The observation of 1.2 mm and 2.0 mm dust continuum emission with NIKA, therefore, provides an excellent opportunity to study variations in the $T_\mathrm{d}$ and $\beta$ on a pixel-by-pixel basis. In addition, the simultaneous acquisition of these images with the same instrument allows many of the systematics to be ignored since atmospheric conditions vary at the same rate in both wavebands, and identical spatial filtering applies in both wavebands once convolved to a common resolution. One can, therefore, reasonably consider that the emission at both wavelengths is spatially coherent (i.e. emitted from a similar volume of space). 

\begin{figure}
	\centering
	\includegraphics[width=\hsize]{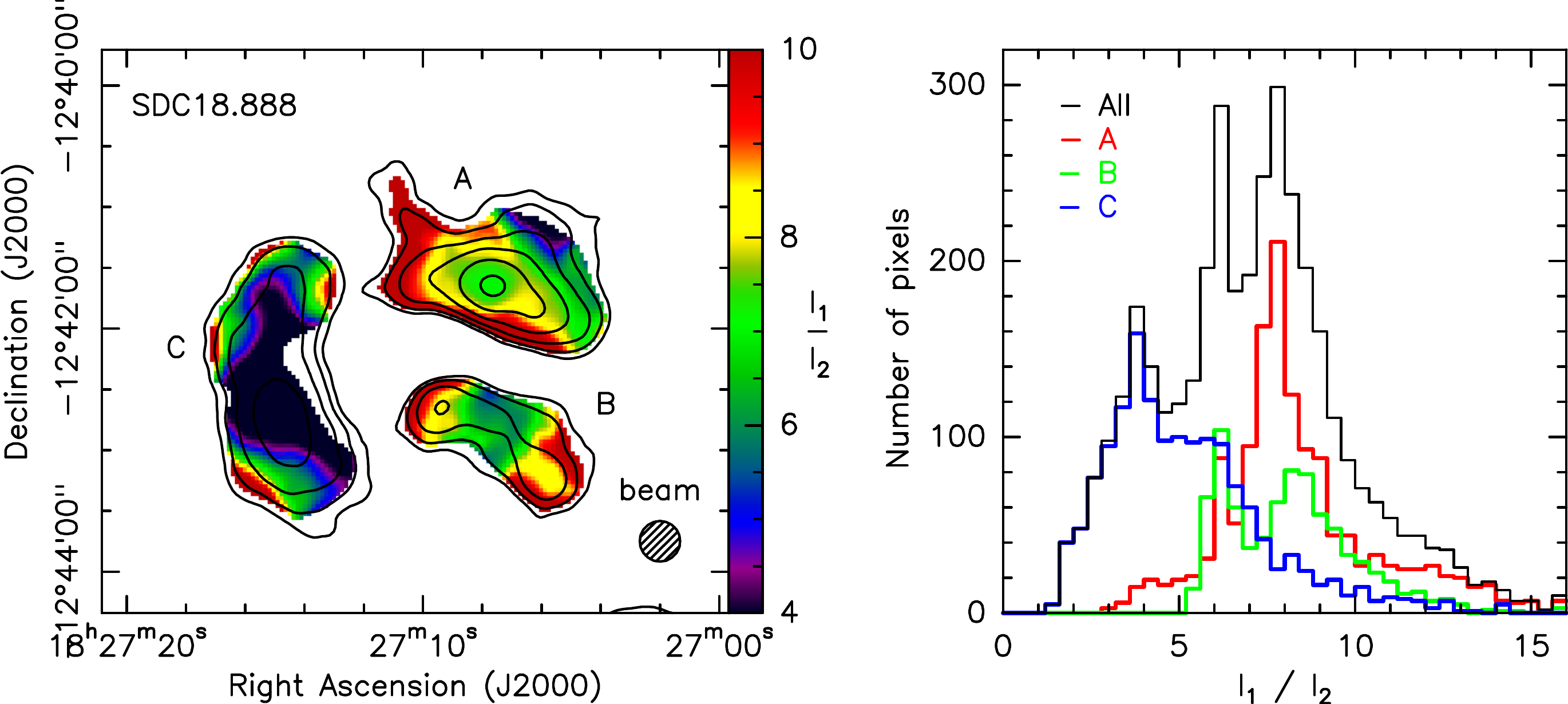}
	\vspace{2.5mm}
	\includegraphics[width=\hsize]{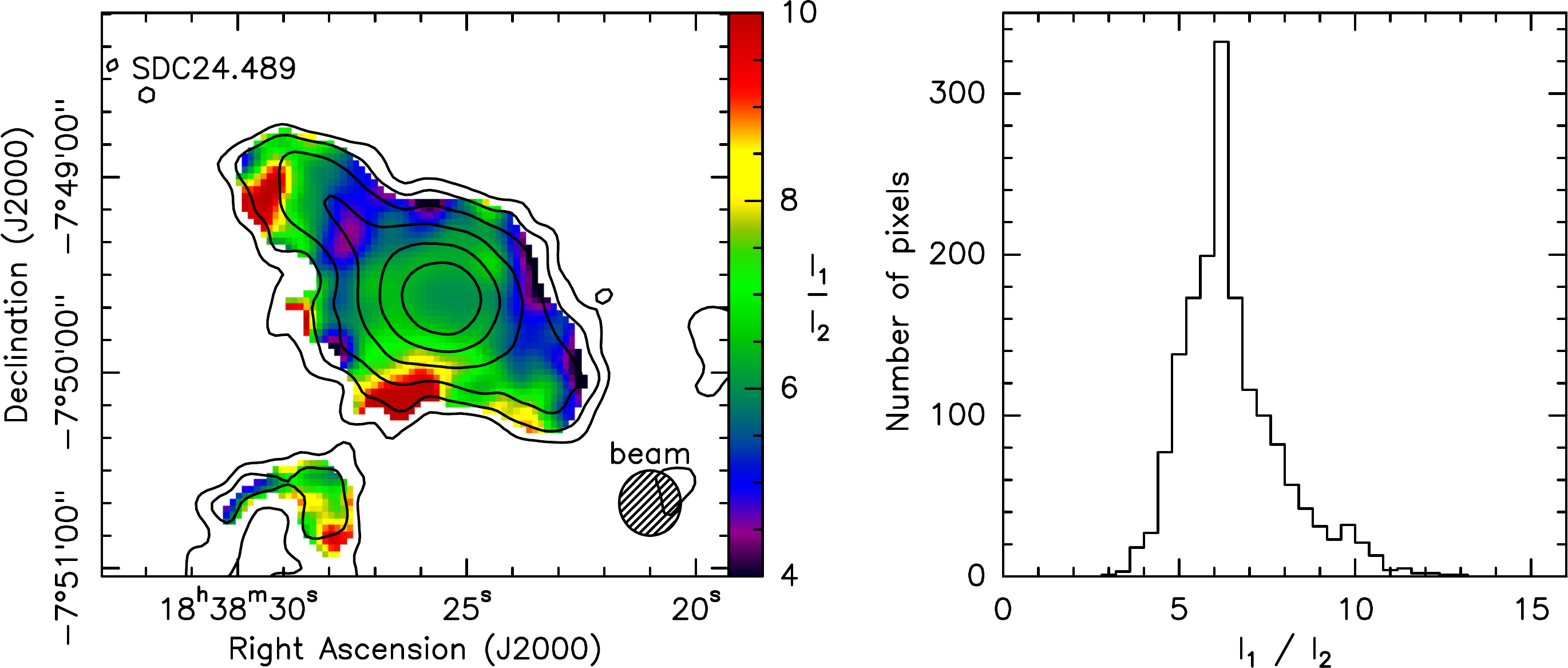}
	\caption{Maps of the 1.2 mm to 2.0 mm intensity ratio for SDC18 (\textit{top}) and SDC24 (\textit{bottom}) at 20\arcsec\ resolution, along with histograms of the pixel values in the regions shown. The contours show the 2.0 mm emission after smoothing to 20\arcsec\ resolution, in terms of the S/N, beginning at the S/N$=2$ level and increasing by a factor of two with each step.}
	\label{fig:ratiomaps}
\end{figure}

Maps of the intensity ratio $I_1/I_2$ were constructed for both IRDCs using the 20 arcsec-resolution images described in Sect. \ref{sec:crkernel}, and are presented in Fig. \ref{fig:ratiomaps} alongside histograms of the pixel values, considering only pixels where S/N$>3$ at both wavelengths and cropping the image to focus on the emission regions. While the pixel values of $I_1/I_2$ in SDC24 are distributed around a single peak at $\sim 6.2$, they are distributed much more widely in SDC18, with multiple peaks at $\sim$4, 6, and 8, roughly corresponding to separate distributions for the three separate structures in the north, south and east which we label A, B and C, respectively. The large variations in SDC18 compared to those in SDC24 may be attributable to astrophysical contamination, which we consider later in this Section.

Before deducing any physical significance of the $I_1/I_2$ maps in terms of $T_\mathrm{d}$ or $\beta$, we should consider all possible synthetic variations acquired during the data collection and processing. Non-physical variations in the ratio maps may arise due to uncertainties in:

\begin{itemize}
	\item \textit{Absolute calibration}: calibration uncertainties were estimated to be 11\% and 9\% for the 1.2 mm and 2.0 mm bands for the NIKA campaign during which these data were taken \citep{Ruppin+17}, and although the calibrations of the wavebands are uncorrelated, the same calibration factor is applied to each pixel for a particular waveband. This would, therefore, introduce a systematic offset, and could not be responsible for introducing any artificial structures.
	\item \textit{Colour correction}: the colour correction applied to the data has a dependency on the source SED. An uncertainty of $\Delta \beta = 0.2$ in the assumed SED provides a fractional uncertainty on the colour correction of  $\lesssim 0.2\%$ which, when compared to the $\sim 10\%$ calibration uncertainty, contributes with a negligible uncertainty to the ratio map.
	\item \textit{Unit conversion}: with different beamsizes in both wavebands, the conversion from the intrinsic units of Jy/beam to the common unit of MJy/sr depends upon the precise beam solid angle. Uncertainties in the beam solid angle therefore translate into uncertainties on the intensity ratio. These uncertainties are accounted for under the figures quoted for absolute calibration.
	\item \textit{Spatial filtering}: the NIKA data reduction pipeline suppresses spatial frequencies on scales above the angular scale of the array footprint, with an identical effect in both wavebands. \citet{Adam+15} show that there is an uncertainty on the flux in each pixel at the 5\% level that is a result of this filtering, though this effect is incorporated into the absolute calibration uncertainty discussed above. To check the validity of this uncertainty in this specific context, we compared ratio maps generated using the 160 \micron\ and 250 \micron\ Hi-GAL images, $R_\mathrm{H} = I_{160}/I_{250}$, and similarly $R_\mathrm{H}^{f}$, which was created using the images spatially filtered in the NIKA manner as described in Sect. \ref{sec:SED}. When we compare the filtered and unfiltered ratio pixel values (after applying our 1.2 mm and 2.0 mm masks), we find that the distribution of the quantity $(R_\mathrm{H}-R_\mathrm{H}^{f})/R_\mathrm{H}^{f}$ has a mean value of $0.0$\% and a standard deviation of 5.8\%, indicating that the 5\% figure is indeed roughly adequate for this analysis.
	\item \textit{Noise effects}: noise arguments provide strong constraints on what can and cannot be trusted in the case of ratio maps. The fractional uncertainty on the intensity ratio is inversely proportional to the S/N, and uncertainties in these data are therefore only low enough to probe dust structure with confidence towards the emission peaks.
	\item \textit{PSF and convolution effects}: as discussed in Sect. \ref{sec:crkernel} the NIKA beam contains a significant fraction of its power in the sidelobes. The power in the sidelobes has been accounted for to a large extent by the use of the convolution kernel described in Sect \ref{sec:crkernel}, though it is still possible that departures from circular symmetry in the beam may introduce artificial signal into ratio maps since they are aligned in the same sense for both wavebands. These artefacts will, however, only apply if the source is sufficiently compact, with high S/N and if only a single scan orientation was used when making the observation. In Fig. \ref{fig:RatioResiduals} we show the residuals after subtracting the ratio map produced using purely Gaussian convolution from the map created using the PSF-matching convolution technique. The exclusive use of Gaussian convolution kernels can introduce ring-like artefacts to the images.
\end{itemize}

\begin{figure}
	\includegraphics[width=\hsize]{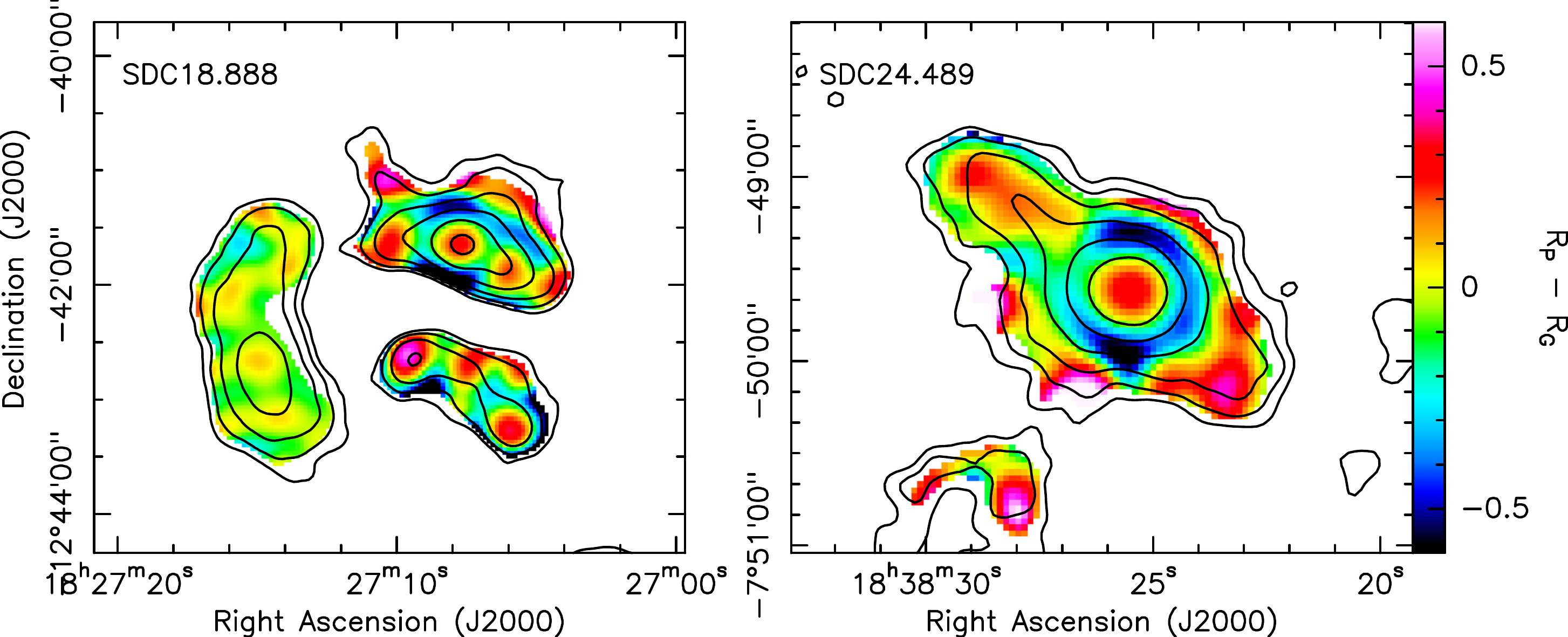}
	\caption{Residuals resulting from the subtraction of the ratio maps convolved using Gaussian kernels alone ($R_\mathrm{G}$) from the ratio maps convolved using the PSF-matching technique ($R_\mathrm{P}$).}
	\label{fig:RatioResiduals}
\end{figure}

The majority of these effects contribute to a systematic offset of every pixel value in the ratio maps, however this is not necessarily the case for noise and PSF effects which may introduce random artefacts into the ratio maps. The impact of the latter two points upon maps of the dust spectral index $\beta$ are explored in further detail in Sect. \ref{sec:betamap}.

In addition, it is possible for the NIKA wavebands to suffer from astrophysical contamination -- that is emission that is unrelated to the thermal continuum emission from dust grains. Inspection of the MAGPIS 20 cm continuum imaging of the two IRDCs reveals that the SDC18 field contains emission that may well be attributable to free--free emission and that might therefore provide a significant level of contamination in the NIKA wavebands. The SDC24 field, by contrast, does not host any detectable 20 cm continuum emission in the MAGPIS image. The 20 cm MAGPIS image of the SDC18 region is displayed in Fig. \ref{fig:20cm} at its native resolution, with the 2.0 mm emission overlaid as black contours. Diffuse emission can be seen across the whole region, but the morphology of the 2.0 mm peak to the east of the image -- region C -- matches the 20 cm continuum emission very closely indeed, indicating a source of contamination that is likely to dominate in that region. The 20 cm emission in regions A and B does not correlate particularly well with the contours of the 2.0 mm NIKA emission, suggesting that they are unrelated and that any free--free contamination in the 2.0 mm band is minimal. We note that in these regions, the 1.2 mm and 2.0 mm morphologies are very similar to each other, which adds weight to an origin in thermal dust emission. There is an exception to the northwest of region A, where 20 cm emission does appear at the edge of the 2.0 mm contours, and the 2.0 mm morphology does differ slightly from the 1.2 mm image.

\begin{figure}
	\centering
	\includegraphics[width=6cm]{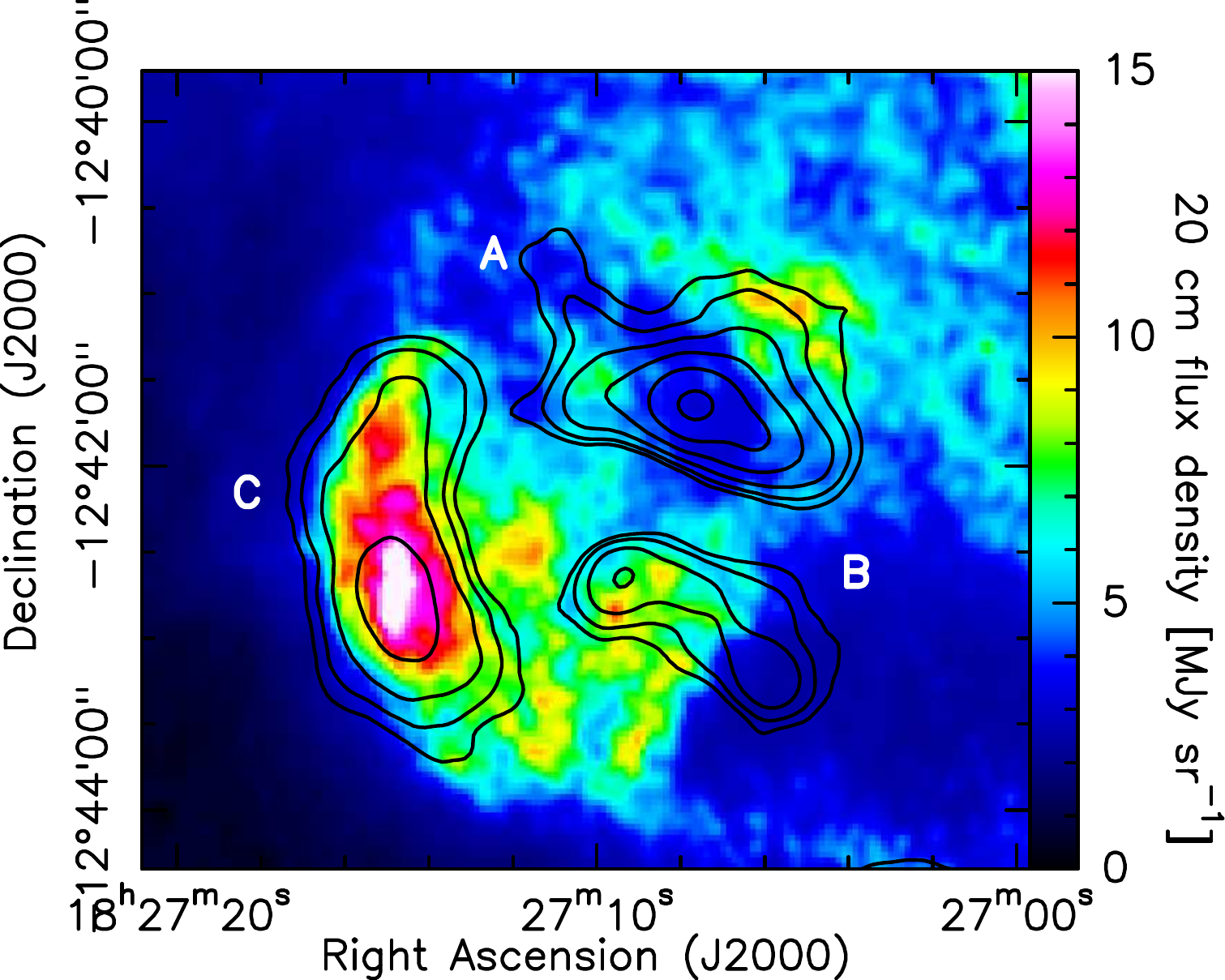}
	\caption{The MAGPIS 20\,cm continuum image of SDC18 at its native 6\arcsec\ resolution. The contours show the 2.0 mm emission as in Fig. \ref{fig:ratiomaps}.}
	\label{fig:20cm}
\end{figure}

We are unable to characterise the SED of the contaminating region (which we explore in more detail in Appendix \ref{sec:contamination}), but in summary we find that a significant proportion of the emission in region C of SDC18 (see Fig. \ref{fig:ratiomaps}) may have arisen from free--free emission associated with the W39 \ion{H}{ii} region, but do not find convincing evidence that such contamination is significant in regions A and B. We also find that contamination from CO (2$-$1) emission is very small, and is present only at the level of $\sim 1-3\%$ across the region, though we are currently unable to measure this at any better than 50\arcsec\ angular resolution. See Appendix \ref{sec:contamination} for more details on our calculation of the CO line contamination.

\subsection{$\beta$ maps} \label{sec:betamap}

The combination of the 1.2 mm and 2.0 mm NIKA data allows us to constrain the dust emissivity spectral index. This can be done by computing the 1.2 mm to 2.0 mm intensity ratio, $I_1/I_2$. This ratio relates to $\beta$ and $T_\mathrm{d}$ through:

\begin{equation}
	\beta= \ln\left(\frac{I_1}{I_2}\frac{B_2(T_\mathrm{d})}{B_1(T_\mathrm{d})}\right)\times \left[\ln\left(\frac{\nu_1}{\nu_2}\right)\right]^{-1},
\end{equation}

\noindent which assumes a single average $\beta$ and $T_\mathrm{d}$ along the line-of-sight. The dust temperatures were determined through the SED fitting of the main clumps -- presented in Sect. \ref{sec:SED} -- and are taken as $T_\mathrm{d} = 14.32 \pm 1.36$ K in SDC18 and $T_\mathrm{d} = 14.15 \pm 1.32$ K in SDC24.  At 1.2 mm and 2.0 mm, $\beta$ only marginally depends on the temperature as the Rayleigh-Jeans regime is approached, meaning that the ratio $I_1/I_2$ is a good proxy for evaluating the variations of $\beta$. However, the assumption of a single dust temperature throughout these IRDCs is simplistic, particularly where embedded objects may be found, and we consider what biases we may thereby introduce later in this Section.

\begin{figure*}
	\centering
	\includegraphics[width=0.9\hsize]{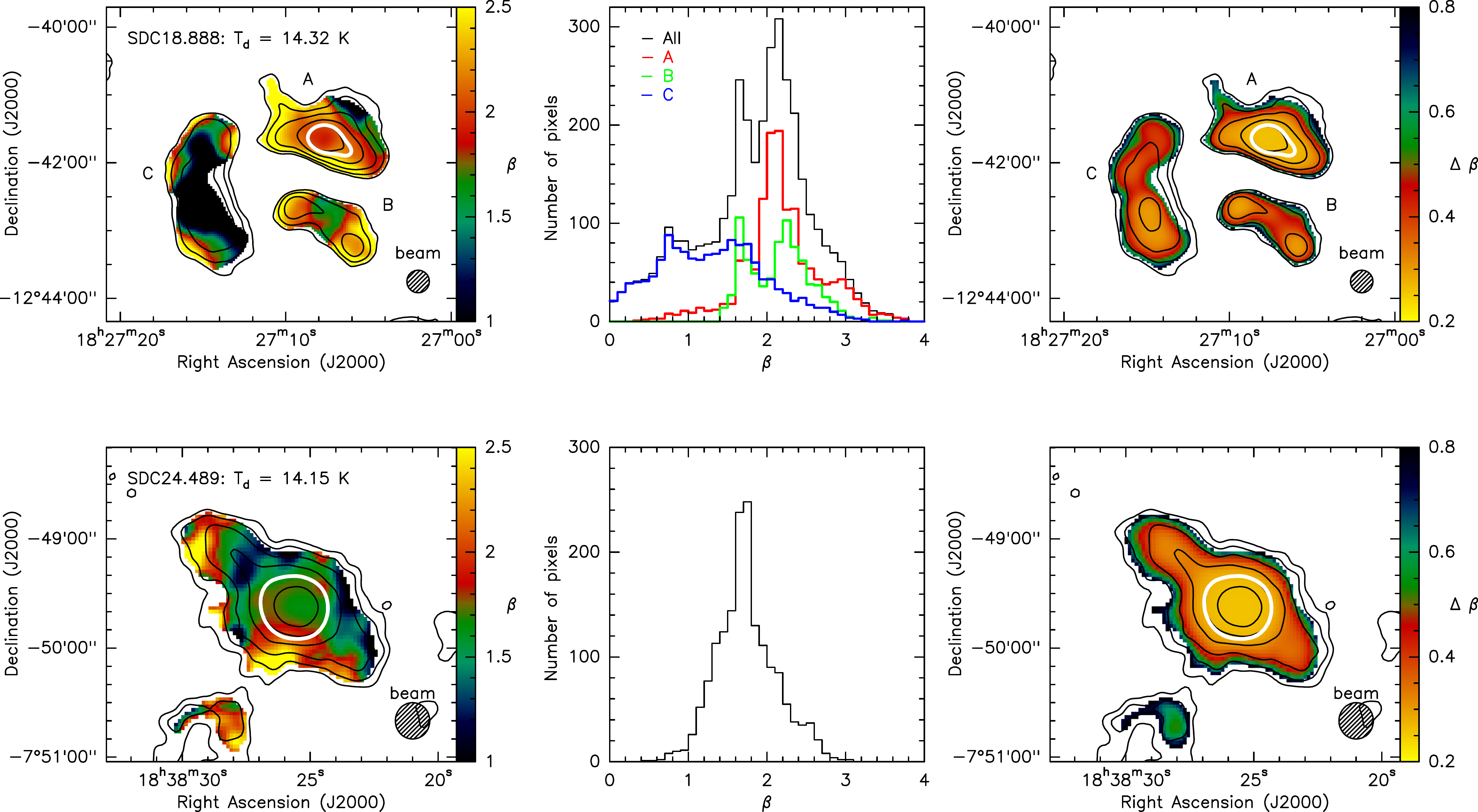}
	\caption{Maps of the dust emissivity spectral index $\beta$ for SDC18 (\textit{top row}) and SDC24 (\textit{bottom row}) with the corresponding histograms in the central column, and the associated total uncertainty maps (including both random and systematic contributions) in the right column. The black contours show the 2.0 mm emission as in Fig. \ref{fig:rawdata}, while the white contours show where the 1.2 mm S/N is equal to 25, above which we can expect the random contribution of the noise to the value of $\beta$ to be $\Delta \beta < 0.1$. The 20\arcsec\ effective beam sizes are shown for scale.}
	\label{fig:betamap}
\end{figure*}

In Fig. \ref{fig:betamap}, maps of $\beta$ are displayed alongside histograms of the pixel values and with the corresponding uncertainty maps. In the case of SDC18, the free--free contamination that afflicted the intensity ratios results in an even greater contrast between high- and low-$\beta$ pixels. The frequency distribution peaks in the $2.1 < \beta \leq 2.2$ bin, with a global mean of 1.83 and a standard deviation of 0.70 though, as was the case with the intensity ratio maps, the three regions contribute different underling peaks to the overall distribution. The distribution spans the range $-0.37 < \beta < 3.50$ though it is likely that much of the low-$\beta$ tail is a result of the free-free contamination which appears to be present in the eastern emission region that lies directly on top of the 20 cm continuum emission region. There appears to be a $\beta$ gradient across the brightest clump (region A) in SDC18, though these appear to be orientated in a similar direction to the 20 cm continuum emission visible in Fig. \ref{fig:20cm}, and consequently may not genuinely be present in that clump. 
   
The SDC24 field, on the other hand, is clear of 20 cm continuum emission and so we can have confidence that free-free contamination is not a significant issue here. The $\beta$ map of SDC24 shows a lot of structure. Most of the extreme values are on the edge of the identified region where the S/N is lowest. The $\beta$ value is approximately 1.65 at the peak of the 2.0 mm emission, and it appears to rise moving out from the centre for a period, before dropping again as the 2.0 mm intensity drops away. This is not azimuthally constant, and there appears to be a difference towards the south of the clump where $\beta$ becomes very high in one particular sector. Variations can also be seen towards the north-east, where a shoulder of 2.0 mm emission between the 3$\sigma$ and 20$\sigma$ levels contains a $\beta$ minimum. The frequency distribution of $\beta$ values peaks at approximately $\sim 1.7$, with a global mean of 1.72 and a standard deviation of 0.34. This is consistent with the value of $\beta = 1.89 \pm 0.18$ derived from the SED fit to the main clump, though the latter represents an average within a 25\arcsec\ radius of its centre.

As an internal consistency check, we compared $\beta$ values derived from our $\beta$ maps to the values recovered from the initial SED-fitting from Sect. \ref{sec:SED} from which we derived our original dust temperatures. In SDC18, using the same 25\arcsec\ aperture as for the SED fit, we measured a mean $\beta$ of 2.23, or a value of 2.11 when each pixel value was weighted by 2.0 mm emission at the same resolution (the latter being more directly comparable to the SED fit value). For SDC24, we obtained a mean aperture value of 1.77 in the non-weighted case, and 1.73 when weighted by the 2.0 mm emission. In both cases, the $\beta$ values derived from the NIKA ratio maps lie within the uncertainty of the values recovered from the SED-fits, though we note that in both cases, the NIKA ratio method recovers lower values.

\citet{Bracco+17} found systematic variations of $\beta$ as a function of radius in the vicinity of two protostellar cores in the Taurus B213 filament, with low values ($\beta \sim 1$) towards the centre, and increasing towards a value at larger radii that is consistent with the measurement in a pre-stellar core in the same filament. We looked for similar systematic variations in SDC18 and SDC24 by constructing radial profiles of $\beta$ using annular apertures on the 20\arcsec-resolution $\beta$ maps of Fig. \ref{fig:betamap}, centred on coordinates of the peak 1.2 emission, which we present in Fig. \ref{fig:betaprofile}. The annuli had a width of 3 arcseconds and measurements were made out to a maximum radius of 40\arcsec\ from the peak of the 1.2 mm emission. The error bars show the mean uncertainty in each annulus resulting from error propagation (effectively aperture photometry on the $\Delta \beta$ map of Fig. \ref{fig:betamap}), while the shaded regions indicate the standard deviations of the pixels in each annulus, disentangling the random and systematic components. The standard deviation in each annulus is the combined effect of both random uncertainties, and intrinsic variations in $\beta$.

\begin{figure*}[t]
	\begin{minipage}{0.5\linewidth}
	\centering
	\includegraphics[width=\textwidth]{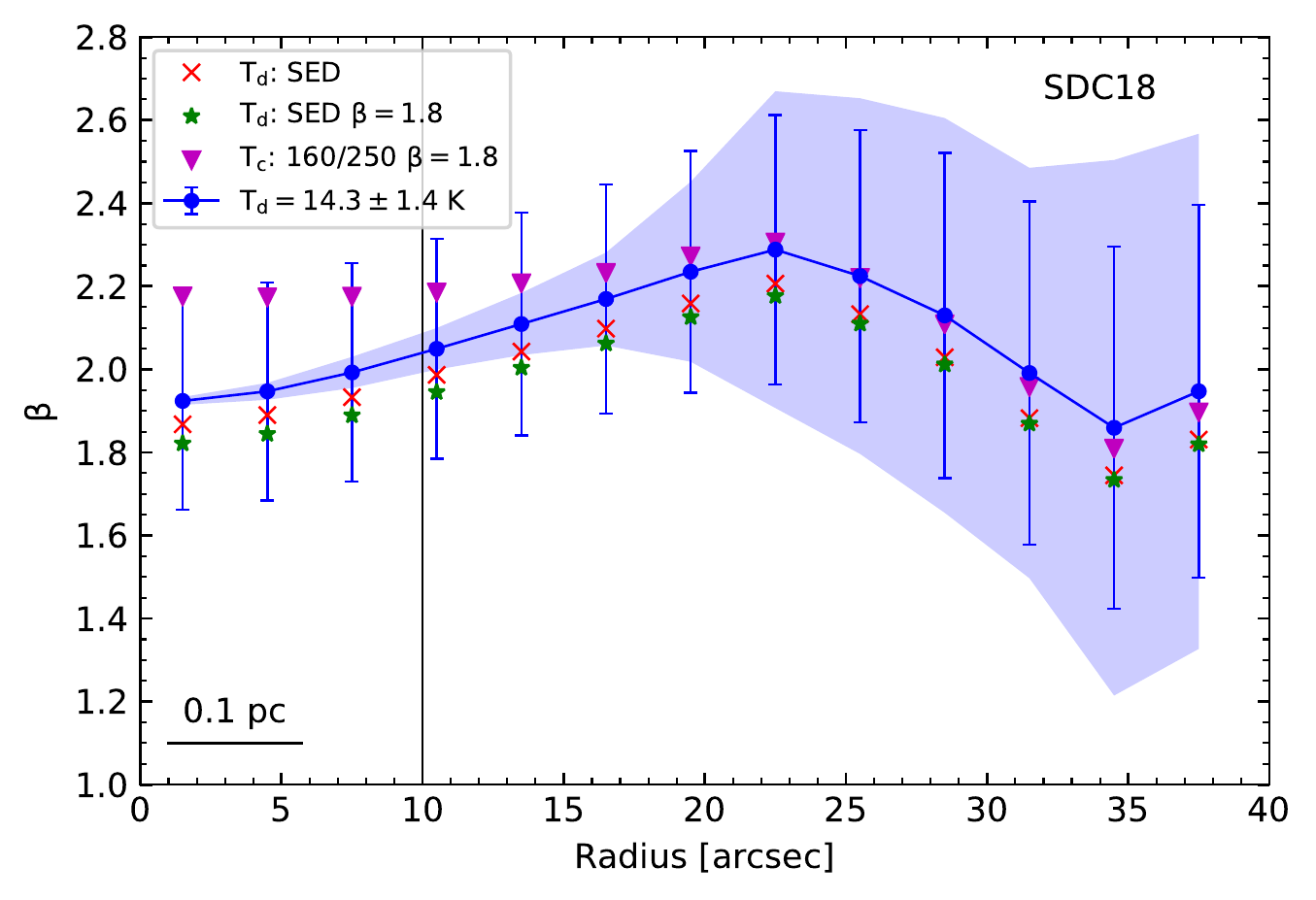}
	\end{minipage}
	\begin{minipage}{0.5\linewidth}
	\centering
	\includegraphics[width=\textwidth]{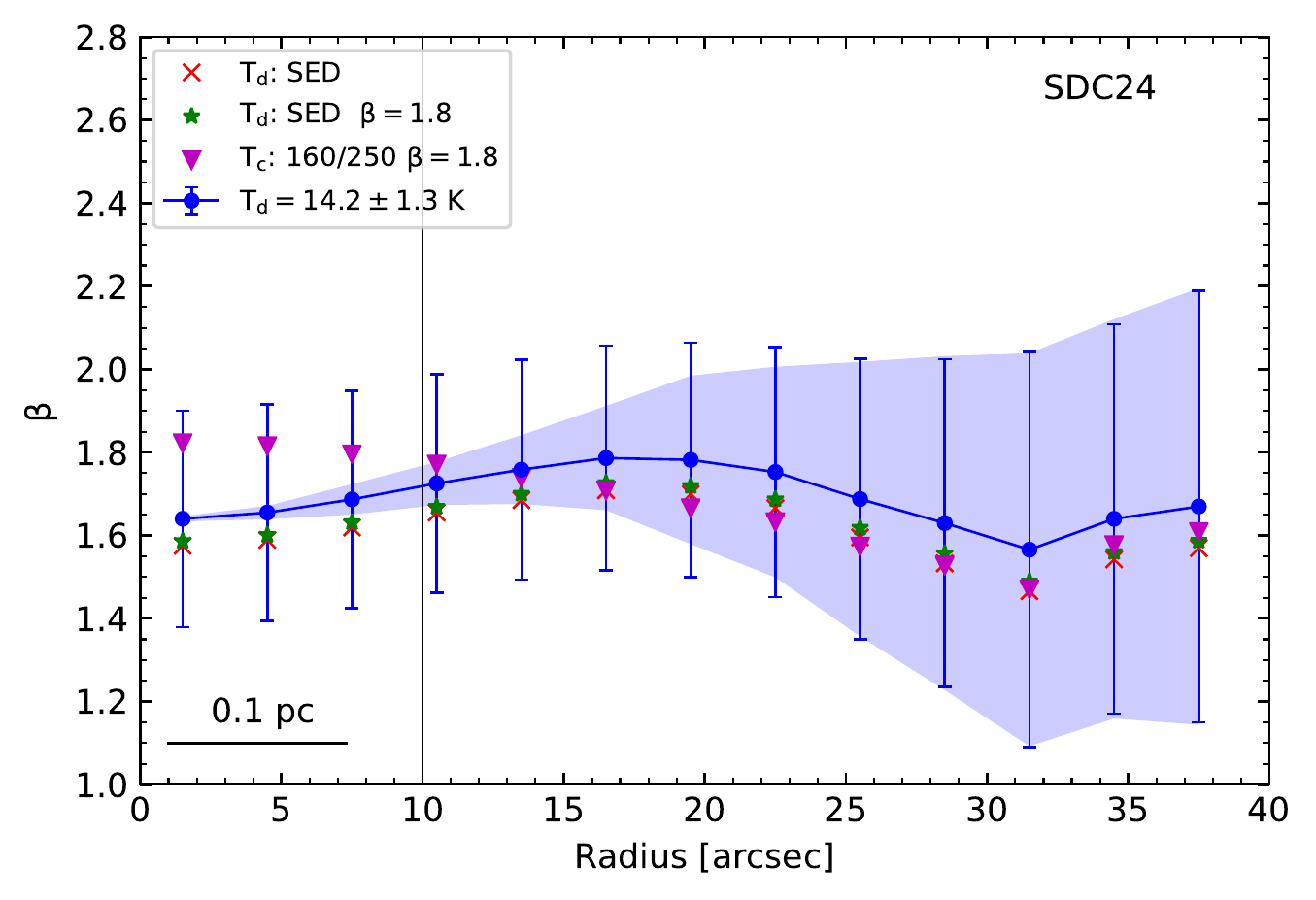}
	\end{minipage}
	\caption{Radial profiles of $\beta$ measured in SDC18 and SDC24. In each case, the blue lines show the mean value of $\beta$ in the annuli centred on each point, and the shaded blue regions show the standard deviation of pixel values within the corresponding annulus. The error bars combine the uncertainties from calibration, noise and temperature uncertainty. The solid black line at a radius of 10 arcseconds shows the extent of the effective beam. The profiles shown as red crosses, green stars and magenta triangles arise from adoption of the different temperature maps used as described in Section \ref{sec:betamap}.}
	\label{fig:betaprofile}
\end{figure*}

In both cases, the radial variations are small when compared to the uncertainties that are dominated at small radii by the calibration uncertainties, while at large radii by noise effects begin to increase the uncertainty. The variation in the outer annuli may also contain genuine variations in dust properties, but we are not sensitive to them using this technique and, in the case of SDC18, while the central $\sim$25--30\arcsec are relatively free from free-free contamination (see Fig. \ref{fig:20cm}), that contamination may have an effect in the outermost annulus. Future studies may improve on this if, for example, Hi-GAL data can be processed through the NIKA pipeline to ensure that equivalent filtering is applied, and variations in $\beta$ could thereby be studied using pixel-by-pixel SED fitting techniques, as used in the JCMT Gould Belt Survey's SCUBA-2 studies \citep[e.g.][]{Sadavoy+13,Chen+16}. The mean values, weighted by $1/\sigma_i^2$, where $\sigma_i$ is the total uncertainty corresponding to each annulus, are $\bar{\beta} = 2.07 \pm 0.09$ and $\bar{\beta} = 1.71 \pm 0.09$ for SDC18 and SDC24, respectively.

In constructing these $\beta$ maps, we made the assumption that a single dust temperature applies over the whole of each IRDC, determined from SED fitting in Sect. \ref{sec:SED}, though this is unlikely to be the case in reality. We calculated additional $\beta$ profiles for the two IRDCs using dust temperature maps generated using several different techniques, which are shown alongside the blue-circle fixed-temperature profile in the bottom panels of Fig. \ref{fig:betaprofile}. The red crosses adopt dust temperatures from maps created using a pixel-by-pixel SED fit using four Hi-GAL images at 160, 250, 350 and 500 \microns. The green stars adopt a temperature map using SED-fitting to Hi-GAL wavelengths again, but this time adopting a fixed $\beta$ of 1.8. Both $\beta$ profiles generated using the temperature maps from Hi-GAL SED fitting show almost identical results, though we note that these temperature maps have an angular resolution of 40\arcsec, a factor of two lower than our NIKA ratio maps. Finally, the magenta triangles use colour temperature maps created following \citet{Peretto+16}, who demonstrated that the 160 \micron\ and 250 \micron\ \Herschel\ imaging can be used to estimate dust temperatures at a resolution of $\lesssim 20\arcsec$. We constructed the 160/250 \micron\ ratio map by using the convolution kernels of \citet{Gordon+08} to account for the non-Gaussian features in PACS and SPIRE beams. This method requires the assumption of a value of the $\beta$, though the apparent circularity in this argument is mitigated by the fact that with the 160 \micron\ to 250 \micron\ intensity ratio, the wavelengths are close to the peak of the dust SED and thus relatively insensitive to $\beta$ in the same way that the 1.2 mm to 2.0 mm intensity ratio -- far out in the tail of the SED approaching the Rayleigh-Jeans limit -- is relatively insensitive to dust temperature. Comparison of profiles made using $\beta=1.6$ to profiles made using $\beta =2.0$ leads to differences in dust temperature of $\sim$1 K. In general, the effect of these varying temperature maps on the $\beta$ profiles is very small, resulting only in small offsets of around $\Delta \beta \lesssim$0.1 -- much smaller than the error bars. The biggest effect arises from the 160/250 \micron\ ratio-derived dust temperatures which are arguably the least robust. In fact, the relative insensitivity of the NIKA intensity ratio-derived $\beta$ map means that only implausibly high dust temperatures could yield a significant deviation.

\section{Mass concentration} \label{sec:Mass}

\subsection{Column density maps}

Column density maps from {\it Herschel} data can be calculated in different ways. The standard way is to perform a four-point (from 160 to 500 \micron) pixel-by-pixel SED fitting. For this, all data must first be smoothed to a common angular resolution of, for example, 40\arcsec, slightly larger than the original 36 \arcsec-resolution 500 \micron\ image to smooth any artefacts resulting from the data reduction procedure as well as noise features. A modified black-body function can then be fitted to the four data points at each position in the map.

As we have seen in the previous Section, $\beta$ does not vary significantly across each IRDC and so we adopted a fixed value for the dust emissivity spectral index of $\beta=2.07$ in SDC18 and $\beta = 1.71$ in SDC24; these values are taken from the weighted mean values and lie within $3 \sigma$ the Galactic average of $\beta = 1.8 \pm 0.2$ as measured by \textit{Planck} \citep{Planck11}. As in Sect. \ref{sec:SED}, we used the \textit{scipy.optimize} $\chi^2$-minimisation routine \textit{curve\_fit} to constrain the two remaining free parameters, $N_{\mathrm{H}_2}$ and $T_\mathrm{d}$, thereby constructing both column density and temperature maps at 40\arcsec\ resolution. The corresponding column density maps are displayed in the top row of Fig. \ref{fig:columndensity}. These images show that the dust emission in both IRDCs is dominated by one source, and all compact sources that can be seen in the NIKA images in Fig. \ref{fig:rawdata} are blurred.

Higher-resolution column density maps can be obtained in several ways by making a few assumptions. The pixel-by-pixel SED-fitting procedure can be repeated with the exclusion of the 500 \micron\ point, carrying out a three point SED fit. Even though the fit is less well-constrained as a result of the decreased number of data points, with two free parameters and therefore one degree of freedom, performing such a fit is still valid, and one can make significant gains in angular resolution, going from 40\arcsec\ to 27\arcsec\ resolution (the raw 350 \microns\ angular resolution is 25\arcsec). Another way to construct column density maps is to calculate a colour temperature using the 160 \microns\ and 250 \microns\ emission, and use Equation \ref{eq:greybody} to combine the resulting temperature map with the 250 \microns\ image to obtain a 20\arcsec\ column density image \citep[e.g.][]{Peretto+16}. In each of these stages, column density maps are obtained using \Herschel\ data alone.

\begin{figure}
	\centering
	\includegraphics[width=\hsize]{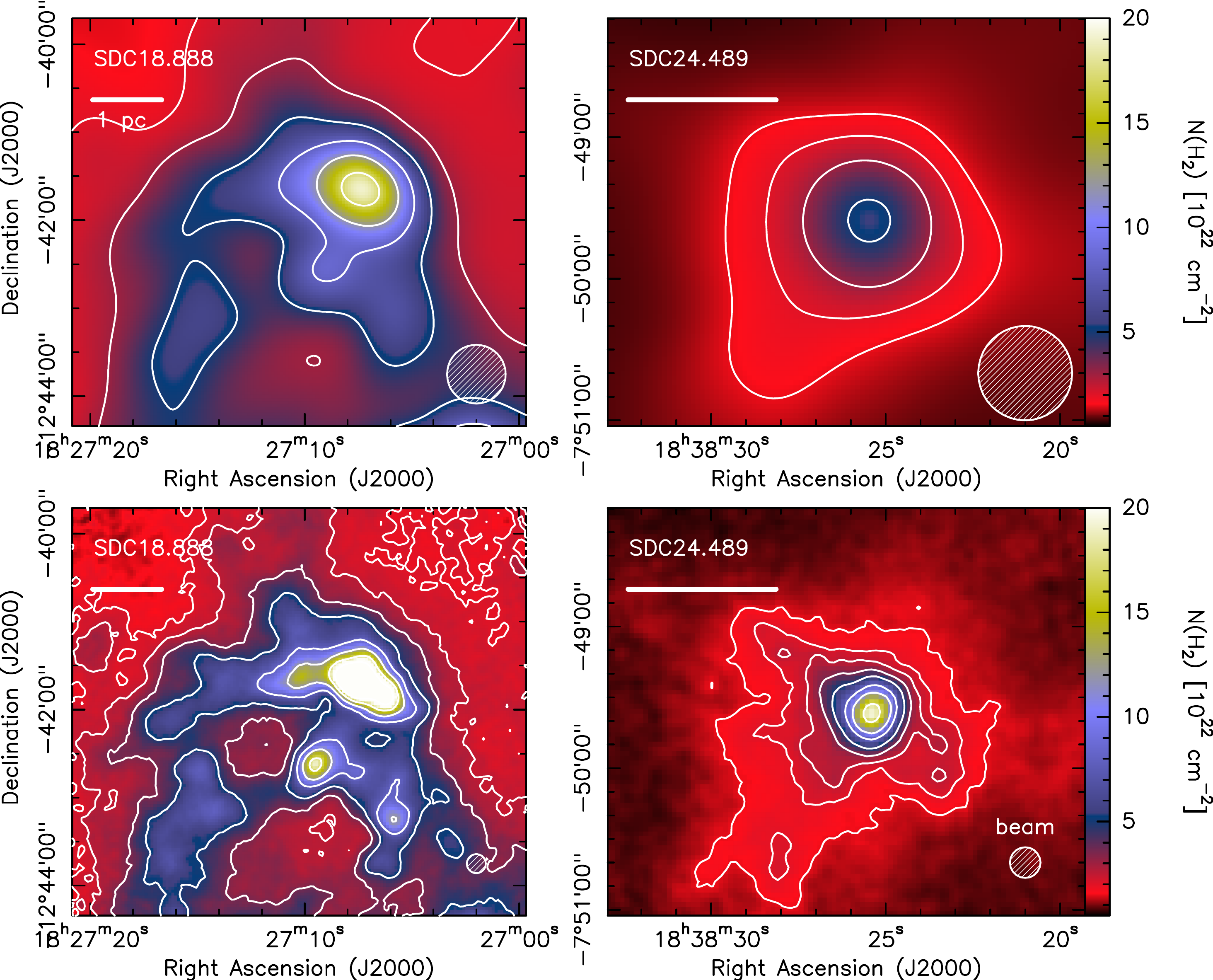}
	\caption{Column density maps for SDC18 (left column) and SDC24 (right column), calculated assuming a single $\beta$ value in each IRDC. The maps on the top row were generated by smoothing Hi-GAL imaging to a common resolution of 40\arcsec, while those on the bottom utilise the multi-resolution technique, sampling between 40\arcsec\ and 13\arcsec\  to combine both \Herschel\ and NIKA data. The SDC18 contours show column densities levels of 2.0, 3.5, 5.0, 8.0, 12.0, 17.0 and 23.0 $\times 10^{22}\,\mathrm{cm}^{-2}$, and the SDC24 are the same with an additional contour at 1.5 $\times 10^{22}\,\mathrm{cm}^{-2}$. Scale bars showing a 1 pc distance are displayed in the top left of each panel.}
	\label{fig:columndensity}
\end{figure}
  
Following \citet{Hill+12} and \citet{Palmeirim+13} it is possible to combine higher-resolution ground-based data with \Herschel\ observations to construct a column density image in which all the information at each wavelength is preserved. The idea is that the four point SED column density map, $ N_{\mathrm{H}_2}^{40\arcsec}$, is the most reliable map for scales beyond 40\arcsec, while the two higher resolution maps, $\widetilde{N}_{\mathrm{H}_2}^{27\arcsec}$ and $\widetilde{N}_{\mathrm{H}_2}^{20\arcsec}$, provide additional information on scale ranges [$40\arcsec-27\arcsec$] and [$27\arcsec-20\arcsec$], respectively. The tilde here signifies that due to the way they have been constructed (i.e. with only a subset of the available data) these column density maps are missing information on large scales. We can construct a fourth column density image, $\widetilde{N}_{\mathrm{H}_2}^{13\arcsec}$, using the NIKA 1.2 mm image in a similar way as for $\widetilde{N}_{\mathrm{H}_2}^{20\arcsec}$, using the same temperature map (at 20\arcsec\ resolution). The following image can then be constructed:

\begin{equation}
 	f_{\mathrm{H}_2}^{l-s}=\widetilde{N}_{\mathrm{H}_2}^s-\widetilde{N}_{\mathrm{H}_2}^s* G_{l},
 \end{equation}
 
\noindent which includes the features on spatial scale range $[l-s]$ to be added to $\widetilde{N}_{\mathrm{H}_2}^{l}$ in order to recover the $\widetilde{N}_{\mathrm{H}_2}^{s}$ column density map. The $G_{l}$ term corresponds to a Gaussian kernel required to smooth the data to an angular resolution $l$. We can finally obtain an expression for a column density map $N_{\mathrm{H}_2}^{13\arcsec}$ at 13\arcsec\ angular resolution that includes all information at all scales using the following equation:

\begin{equation} \label{eq:multires}
	N_{\mathrm{H}_2}^{13\arcsec}= N_{\mathrm{H}_2}^{40\arcsec} +  f_{\mathrm{H}_2}^{40\arcsec-27\arcsec}+  f_{\mathrm{H}_2}^{27\arcsec-20\arcsec} + f_{\mathrm{H}_2}^{20\arcsec-13\arcsec}.
\end{equation}

The fact that spatial frequencies on arcminute scales have been filtered out of the NIKA data does not matter since we only retain features between 13\arcsec\ and 20\arcsec\ of the $\widetilde{N}_{\mathrm{H}_2}^{13\arcsec}$ column density image. We produced 13\arcsec\ column density maps in this way, and the resulting images are displayed in the bottom row of Fig. \ref{fig:columndensity}. We can see that we have recovered many more structures compared to the 40\arcsec\ column density map. In particular, we can see the presence of several cores in the SDC18 that are not discernible in the 40\arcsec\ map, while SDC24 remains dominated by a single clump. Filamentary structures can be seen in SDC18 that correspond with features in the 8 \micron\ absorption of Fig. \ref{fig:rawdata}, though the same can not be said for SDC24 for which the finer structures remain unresolved. We remind the reader that for SDC18, the structure to the east of the image at least (referred to as region C in Figs. \ref{fig:ratiomaps} and \ref{fig:betamap}), is suspected to have arisen from free-free contamination, and is probably not an authentic dust structure.

The various assumptions required to produce the 13\arcsec-resolution column density map, $N_{\mathrm{H}_2}^{13\arcsec}$, means that it is intrinsically less accurate than the column density map $N_{\mathrm{H}_2}^{40\arcsec}$. To test the magnitude of this difference, we smoothed the $N_{\mathrm{H}_2}^{13\arcsec}$ using a Gaussian kernel of FWHM 37.8\arcsec\ back to the original 40\arcsec, and construct maps of the ratio of the smoothed $N_{\mathrm{H}_2}^{13\arcsec}$ to the $N_{\mathrm{H}_2}^{40\arcsec}$ map. The distributions of pixel values in these column density ratio maps for SDC18 and SDC24, peak at 1.002 and 1.003 and with standard deviations of 0.006 and 0.007, respectively. The high-resolution column density maps contain 0.2$-$0.3\% more mass in total, when compared to the more robust low-resolution maps, and pixel-to-pixel variations are on the order of 0.5\%, giving us a high level of confidence in this methodology. In the analysis that follows from the combined $N_{\mathrm{H}_2}^{13\arcsec}$ map, we have adopted uncertainties derived from a corresponding error map, in which the uncertainties arising from the absolute calibration of the \textit{Herschel} and NIKA data have been propagated through each of the constituent stages given in Eq. \ref{eq:multires}.

\subsection{Mass determination}

The hierarchies of substructures in the multi-resolution column density maps of the two IRDCs were analysed by using the Python \textit{astrodendro} implementation of the dendrogram algorithm \citep{Rosolowsky+08}. Dendrograms segment complex structures as a function of contour level, describing structures as `branches' and `leaves', where branches are structures that contain further substructures within them and leaves contain no further substructure, thus representing the most fundamental units in the hierarchy. The algorithm is sensitive to only three parameters: a zero-level contour, a minimum intensity `dip' between separate structures, and a minimum number of pixels required. We summarise our choice of values for these parameters in Table \ref{tab:dendrogram}. We defined the zero-level contour separately for each image, based on the extent of the structures visible in each frame. The Hi-GAL images, used in the construction of the most extended column density map components, are limited by cirrus noise as opposed to photon noise, and the S/N is high everywhere. Consequently, defining the zero-level based on a detection limit would have little utility. Identical parameters for the minimum dip and the minimum number of pixels required for each clump were used for both regions to produce the most consistent dendrogram segmentations for the images with the same angular resolution. The minimum number of pixels was determined by the area of the highest-resolution component of the column density map, with a 13\arcsec\ effective beam size.

We chose contour levels identified by the dendrograms that isolated the cloud structures in which we are interested, and excluded structures outside of those boundaries. The region edges are defined by contour levels of $3.5 \times 10^{22}$ \pcmm\ in SDC18 and $1.5 \times 10^{22}$ \pcmm\ in SDC24, and we defined these levels as the lower limit for column density per pixel to be included in the mass estimates (i.e. the background level). In the following text we adopt the terminology of \citet[][see Figure. 4 of that study]{Rosolowsky+08} who describe a number of methods for interpreting isosurfaces of emission in contour-segmented maps, such as our column density maps. The so-called `bijection' method simply sums all pixel values within the boundary (in this case, a column density contour) defining an object. Here, this represents an upper limit to the mass of a cloud that is sitting a diffuse lower-column density envelope by incorporating that diffuse column density into its mass. In the `clipping' technique, a background subtraction is carried out, subtracting all of the intensity (i.e. mass) that falls below the contour level at the boundary, and thus defining a lower limit of the mass. Since we will discuss structures within an IRDC, all measurements in this Section refer to clipped masses, unless otherwise specified. We note that by construction these mass estimates are identical at the upper size limit of the cloud boundary.

\begin{table}
	\caption{Parameters used in the \textit{astrodendro} analysis.}
	\label{tab:dendrogram}
	\centering
	\begin{tabular}{c c c c}
	\hline\hline 
	Source		& \texttt{min\_value} & \texttt{min\_delta}	& \texttt{min\_npix}  \\
	\hline \\[-2ex] 
   	SDC18 		& $3.5 \times 10^{22}$ \pcmm 	& $0.1 \times 10^{22}$ \pcmm & 24 \\
	SDC24 		& $1.5 \times 10^{22}$ \pcmm 	& $0.1 \times 10^{22}$ \pcmm & 24 \\
	\hline 
	\end{tabular}
\end{table}

\begin{figure}
	\centering
	\includegraphics[width=\hsize]{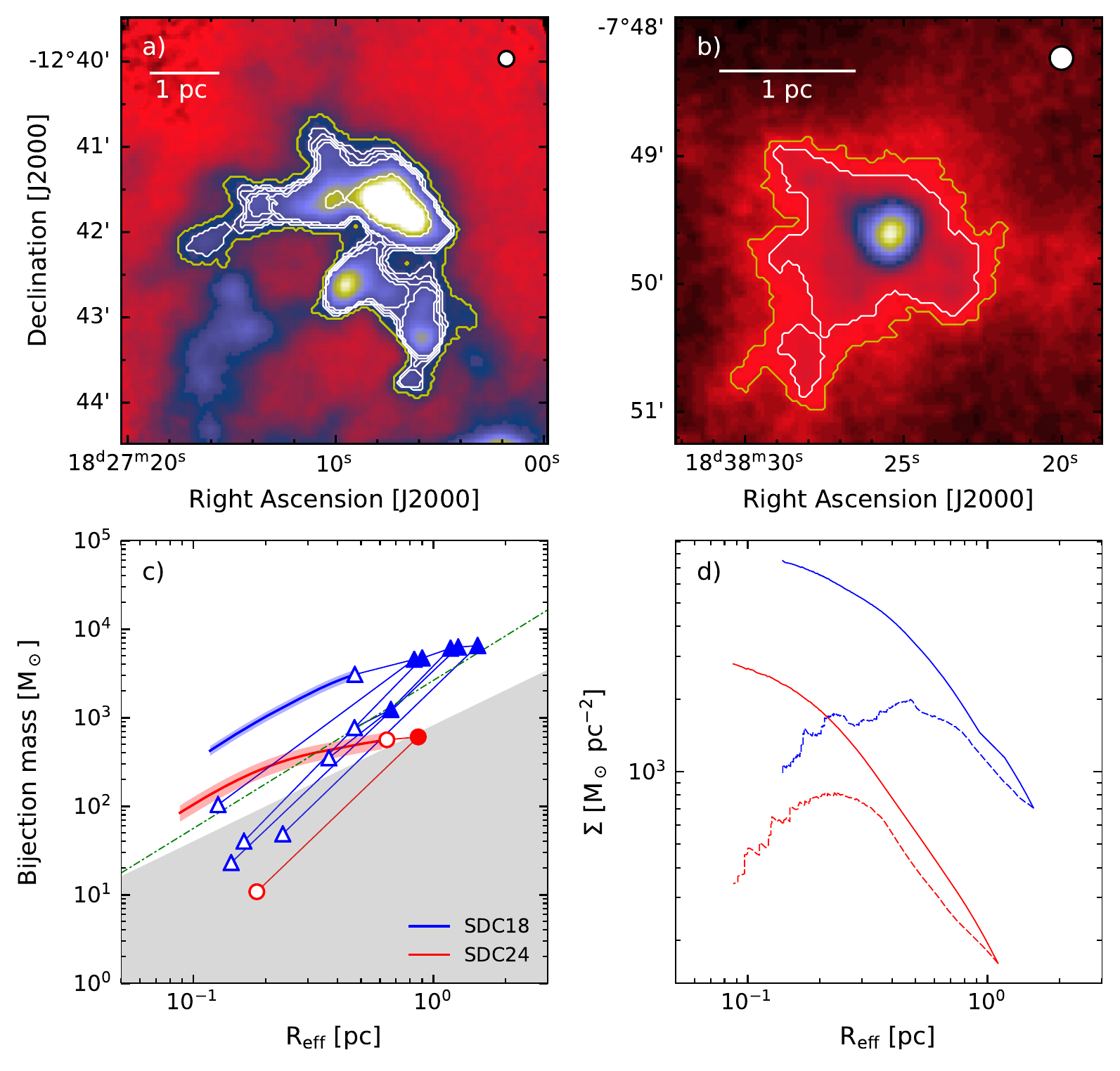}
	\caption{Panels a) and b) show the contours that determine the boundaries of the dendrogram substructures within SDC18 and SDC24, respectively, with leaves and branches shown in white and the IRDC boundaries shown in yellow. The underlying image shows the combined-resolution column density map from Fig. \ref{fig:columndensity}. Panel c) shows the enclosed (bijection) masses of the dendrogram substructures as a function of the effective radius, $R_\mathrm{eff} = \sqrt{A/\pi}$, for SDC18 in blue triangles and SDC24 in red circles with filled points showing the dendrogram `branches' and empty points corresponding to the `leaves'. The solid red and blue lines connect each substructure to its parent, where applicable. The shaded region defines objects that do not satisfy the \citet{Kauffmann+Pillai10} criterion for HMSF clumps, $m(r) \geq 870 \, \mathmsol \, (r/\mathrm{pc})^{1.33}$. The dot-dashed green line shows the mass-radius relation calculated for ATLASGAL clumps that exhibit signs of ongoing high-mass star formation \citep{Urquhart+14} with $m(r) = 2630 \, \mathmsol \, (r/\mathrm{pc})^{1.67}$. The solid blue and red lines show the mass distributions inside the boundary of the most massive leaf in each IRDC, with uncertainties shown by the corresponding shaded areas. Panel d) shows the mean surface density as a function of radius for all contour levels in each cloud that lead to the most massive clumps. The solid and dashed lines represent upper limits, calculated from the bijection and clipping methods, respectively.}
	\label{fig:massconcentrationA}
\end{figure}

In Fig. \ref{fig:massconcentrationA} we present results based on the dendrogram structures. Panels a) and b) show the multi-resolution column density map in greyscale with the contour levels that define the different structures overlaid. In both cases, the yellow contour level is that selected to define the boundary of the dust structures though, while in SDC24 this corresponds to the lowest level in the dendrogram, in SDC18 it is the first level that excludes the free-free contaminant to the south-east. SDC18's boundary is defined by a contour level of $4.7\times 10^{22}$ \pcmm, which encloses a mass of $6510 \pm 800$ \msol\ within an effective radius of 1.53 pc. The boundary of SDC24 has a contour level of $1.5 \times 10^{22}$ \pcmm, enclosing a mass of $610 \pm 130$ \msol\ within a 0.87 pc effective radius, an order of magnitude less massive than SDC18. Enclosed contours that contain no further substructures are the leaves, of which there are seven in SDC18 and two in SDC24. Panel c) of Fig. \ref{fig:massconcentrationA} shows the enclosed mass (calculated using the bijective method) as a function of the effective radius, $R_\mathrm{eff} = \sqrt{A/\pi}$, which is the radius of a circle with the equivalent area, $A$, of each structure. The filled and empty triangles and circles show these quantities for all structures and the leaves, respectively, and the solid blue and red lines follow the mass distribution within the most massive leaf (i.e. dense cores) of each region, with uncertainties given by the shaded regions.

We compare the mass-radius relationship for the structures in Fig. \ref{fig:massconcentrationA} panel c) with the \citet{Kauffmann+Pillai10} limit for HMSF with the shaded region denoting the region in which clouds or clumps would not be expected to go on to form high-mass stars. The dominant clumps in both IRDCs lie in the non-shaded region, satisfying this criterion for HMSF. We also note that these mass profiles for both clouds seem to follow broken power laws (in the case of SDC18, the profile can be traced back through its parent structures which are the solid triangles at larger effective radii), indicating a change in the volume density profile of these IRDCs -- becoming flatter in the inner region, at $R_\mathrm{eff} \gtrsim 0.5$ pc in SDC18 and $R_\mathrm{eff} \gtrsim 0.3$ pc in SDC24. We note that another three of the dendrogram leaves of SDC18 also fall above the \citet{Kauffmann+Pillai10} relation, though we stress that we therefore interpret these as HMSF `candidates' since such a criterion is not conclusive on its own.

In panel d) of Fig. \ref{fig:massconcentrationA}, the surface density is explored as a function of radius within the most massive clumps and we see a turnover at an effective radius of $\sim 0.5$ pc in SDC18 and $\sim 0.3$ pc in SDC24. Outside of these radii, the gradients of the upper (solid, bijection) and lower (dashed, clipping) limits converge, but the picture gets more confusing inside. In both cases, the mid-point of the limits would appear to more-or-less flatten off, indicating that the volumetric density is behaving approximately as $\rho(r) \propto r^{-1}$ in the inner part of the clouds, and falls off more rapidly at larger radii. 

\begin{figure}
	\centering
	\includegraphics[width=\hsize]{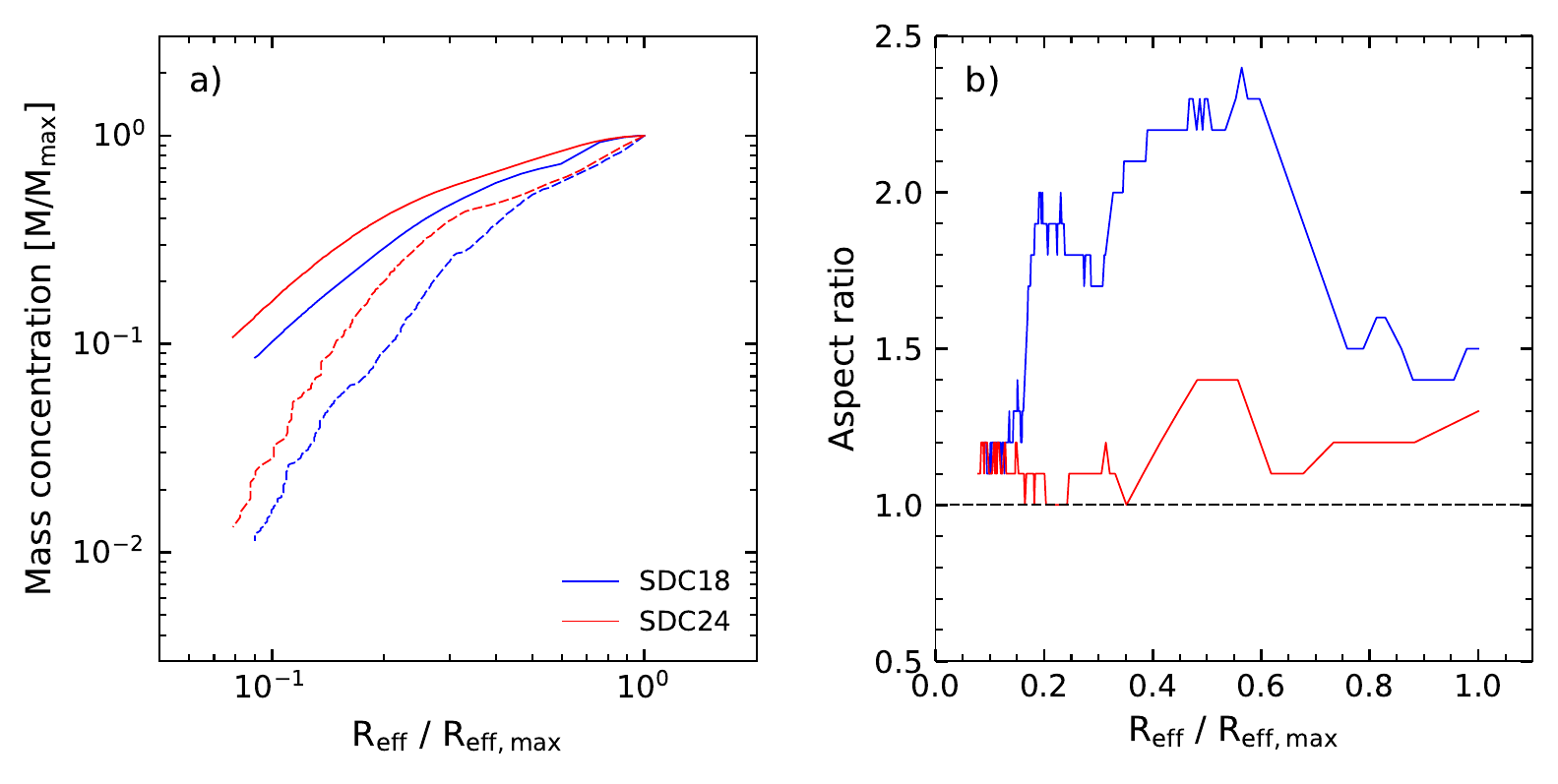}
	\caption{a) the concentration of mass as a function of radius within the two IRDCs. The solid and dashed lines show the upper and lower limits on the mass, respectively, calculated as in panel d) of Fig. \ref{fig:massconcentrationA}. b) the aspect ratio (as defined by \citealt{Peretto+Fuller09}) of each structure as a function of the normalised effective radius.}
	\label{fig:massconcentrationB}
\end{figure}

Following up on our discussion of the steepness of the density profiles, we decided to compare the mass concentration of both IRDCs. Fig. \ref{fig:massconcentrationB} a) shows the normalised mass profiles for both IRDCs (normalised both in radius and mass) starting from the lowest level in the IRDCs -- the yellow boundaries in panels a) and b) of Fig.~\ref{fig:massconcentrationA} -- and following all substructures leading to the most massive clump of each IRDC. At contour levels higher than the boundaries of the leaves, the dashed profiles become equivalent to the shaded solid lines in panel c) of Fig.~\ref{fig:massconcentrationA}. In panel a) of Fig.~\ref{fig:massconcentrationB} we see that, despite being an order of magnitude less massive, SDC24 is about a factor of $\sim 2$ more concentrated than SDC18 at most (normalised) radii. We calculated the aspect ratio for each contour level, following the method of \citet[][see their Appendix A for a full formulation]{Peretto+Fuller09}, which takes the ratio of the mass-weighted rms extent in the direction each of the major and minor principal axes of inertia. In panel b) of Fig.~\ref{fig:massconcentrationB} we can see that, at each level, the mass distribution in SDC24 is much more circular (aspect ratio close to 1), and SDC18 only converges to an aspect ratio of $\sim 1$ as the contour level reaches a single beam in size. Since these aspect ratios were estimated from the 2D projections of these objects, these are lower limits on the 3D aspect ratios of these clouds, and the difference between the two IRDCs could potentially be much larger than a factor of 2.

\section{Discussion} \label{sec:Discussion}

\subsection{Evolution of dust properties}

The evolution of dust properties and the emissivity spectral index $\beta$ is a long-standing issue. \citet{Sadavoy+13} created maps of $\beta$ by using pixel-by-pixel SED fitting technique to combine \Herschel\ PACS and SPIRE data at 160--500 \micron\ with JCMT SCUBA-2 450 and 850 \micron\ data, finding that the addition of the long-wavelength SCUBA-2 data allows significant improvements in the determinations of $\beta$ and dust temperature to be made. Variations have been observed in various different datasets, and all found an anti-correlation between dust temperature and $\beta$. The so-called $T_\mathrm{d}-\beta$ degeneracy can result from noise when using $\chi^2$-minimisation SED-fitting methods, though this effect should be minimised by reducing datasets to include only areas with very high S/N. Despite this $T_\mathrm{d}-\beta$ degeneracy, several studies have concluded that the measurement degeneracies are not sufficient to account for the entirety of the anti-correlation \citep[e.g.][]{Planck11,Chen+16}, and an intrinsic physical anti-correlation has also been measured in laboratory experiments \citep[e.g.][]{Agladze+96}.

We see no evidence of significant systematic radial variations in the dust emissivity spectral index, $\beta$, within the two IRDCs in our maps generated from the ratio of the two NIKA wavebands, regardless of the dust temperature model we adopt. However, we do see a significant difference in the $\beta$ value between the clouds. Since these two images were taken in the same night, we can say that the absolute calibration is the same in both maps, and so they should have the same systematic offset. Taking account of this, we can say that the weighted mean $\beta$ values of $\bar{\beta} = 2.07 \pm 0.09$ in SDC18 and $\bar{\beta} = 1.71 \pm 0.09$ in SDC24 are significantly different, indicating that the different environments are  affecting the dust properties on spatial scales of $\sim 1$ \ pc at least. The uncertainty in the absolute calibration of both maps prevents us from telling whether $\beta$ is being elevated relative to the Galactic average of $\beta \approx 1.8$ in one cloud or whether it is being suppressed in the other. Higher-than-average $\beta$ values in SDC18 could be explained by its location in the vicinity of the W39 \ion{H}{ii} region. We note that although we would expect raised dust temperatures in such an environment, we recover similar temperatures in SDC18 to the quiescent environment of SDC24. The $T_\mathrm{d}$$-$$\beta$ degeneracy is insufficient to explain this, as repeating the SED fitting for SDC18 with dust temperatures raised to 17 K or 20 K yields poor results. The raised $\beta$ is, perhaps, best explained by shocked regions along the line of sight being incorporated into the column-averaged measurement.

Since we cannot assume a correlation between the calibration of NIKA wavebands, we cannot constrain the absolute value of $\beta$ any more accurately than to within $\Delta \beta \approx 0.25$ with NIKA data alone (a figure which arises directly from the 11\% and 9\% uncertainties in $I_1$ and $I_2$), even with extremely good S/N and the negligible uncertainties on the dust temperature at these wavelengths. Calibration of the background levels of the NIKA data using the low-resolution but space-based \textit{Planck} observations provides a feasible route to calibrate NIKA maps absolutely. Relative differences between $\beta$ values can still be measured, though the search for radial trends in $\beta$ should, in future, be carried out on maps with either higher sensitivity or brighter sources than we have in this study. A relative uncertainty in the NIKA ratio-derived $\beta$ maps of $\Delta \beta < 0.1$ can be achieved over regions in which we can achieve S/N\,$\gtrsim 25$. Higher sensitivity observations over larger areas using this technique do have the potential to allow a study of systematic variations as a function of environment.

Recent \textit{Planck} results do, however, allow use to place our $\beta$ values into some additional context, and we compare our results to the latest all-sky dust temperature and $\beta$ maps derived by \citet{PlanckIntXLVIII}. The effective resolution of \textit{Planck} is much coarser than NIKA in these maps, with a 5.0\arcmin\ FWHM, and so neither IRDC is resolved. For SDC18, the \textit{Planck} map has a value of $T_\mathrm{d} = 22.3 \pm 1.1$ K and $\beta = 1.82 \pm 0.09$, at the location of the 1.2 mm emission peak, and for SDC24 values of $T_\mathrm{d} = 21.1 \pm 1.6$ K and $\beta = 1.78 \pm 0.15$ are recovered. In both cases, the NIKA data recover dust temperatures that are roughly 8 K lower than those derived from the \textit{Planck} data, though the $\beta$ values fall within the quoted total uncertainties. The fact that we recover systematically lower temperatures for our IRDCs is not surprising; the spatial filtering applied to ground-based data such as our NIKA images biases the emission towards compact sources, and compact infrared-dark sources will also be cooler than the surrounding ISM. In addition to the effect of the spatial filtering is the relative beam dilution, which is significant when comparing 20\arcsec-resolution data to that at $\sim$5\arcmin, and this effect is compounded by further line-of-sight averaging.

While we do see a marginal ($\Delta \beta \sim \,$0.1$-$0.3) increase in the mean value of $\beta$ out to a radius of $\sim$20\arcsec\ in both IRDCs,  the magnitude of this increase is small when compared to the variations in terms of the standard deviation of $\beta$ values as a function of radius. The error bars in Fig. \ref{fig:betaprofile} within a radius of $\sim$30\arcsec\ are dominated by a systematic calibration offset, whereas the random uncertainties are better represented by the shaded regions that show the 1$\sigma$ deviations from the mean (though this probably also includes intrinsic variations in $\beta$) and so, while all of the radial variations we see are consistent with a constant central value, we are still sensitive to relative radial variations. However, beyond $\sim$30\arcsec, noise effects begin to dominate as the S/N becomes low, and the visible relative differences can not be trusted. The $\beta$ structure away from the emission peak in SDC24 visible in Fig. \ref{fig:betamap} is not azimuthally constant about the emission peak, and we explore how this structure might arise artificially in Appendix \ref{sec:syntheticbetamaps}. The locations of artificial structures arising through noise effects should correlate with low S/N regions, which is not the case for all of the variations in this IRDC, though we cannot exclude the possibility that some of these structures have arisen through beam asymmetries. The effects of such asymmetries are difficult to model, but their impact can be reduced when making the observations by observing in such a way that the position angle of the beam pattern varies with respect to the source.

\begin{figure}
	\centering
    \includegraphics[width=0.6\hsize]{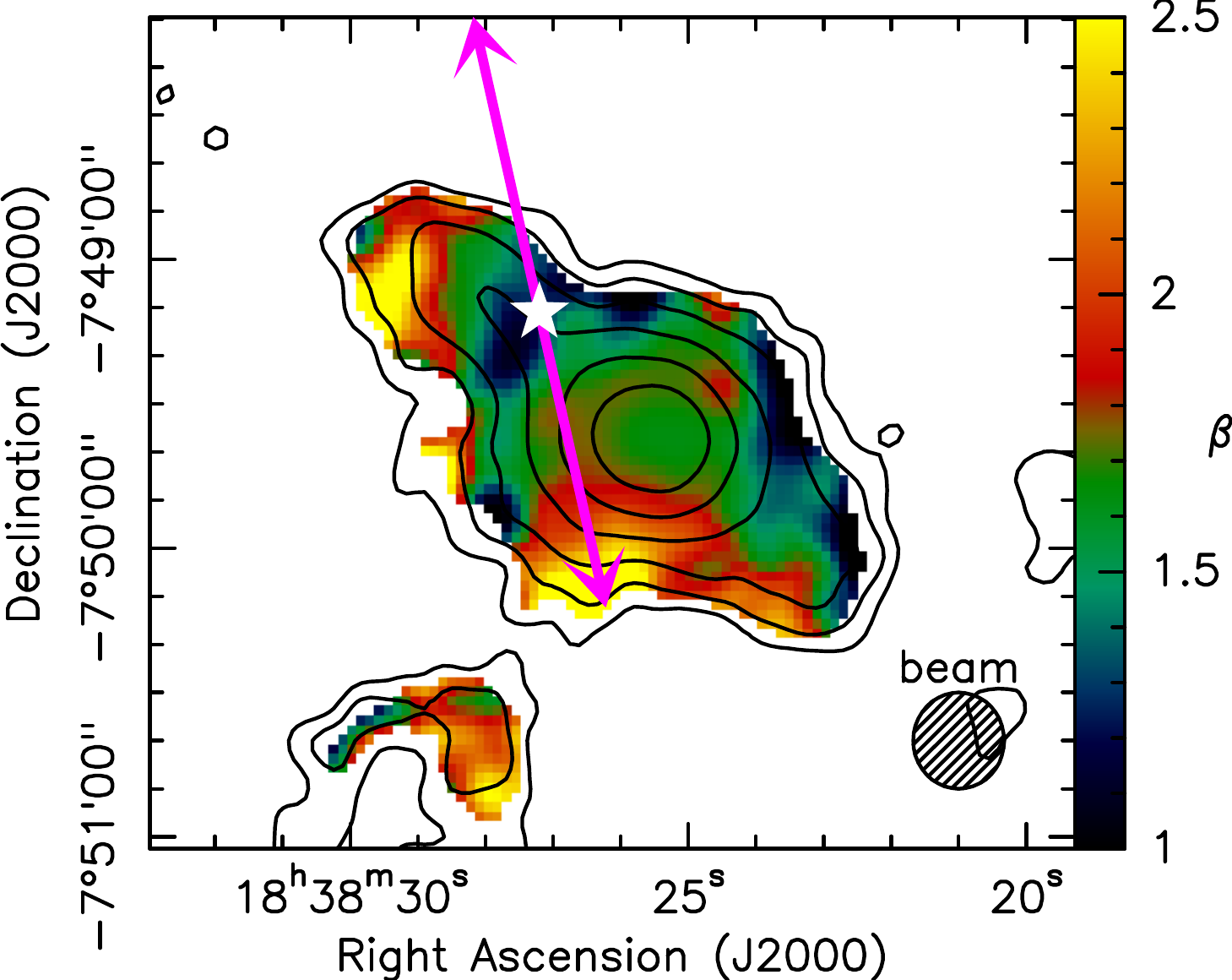}
    \caption{$\beta$ map for SDC24 overlaid with the location of MHO 3240 -- the driving source candidate (white star) -- associated with the molecular hydrogen outflows (magenta arrows) identified by \citet{Ionnidis+Froebrich12}. The contours show the 2.0 mm S/N as in Fig. \ref{fig:ratiomaps}.} 
    \label{fig:outflows}
\end{figure}

The SDC24 region hosts a bipolar molecular hydrogen outflow (MHO) identified by \citet{Ionnidis+Froebrich12} using observations of the 1$-$0 S(1) line of H$_2$ at 2.122 \micron\ from the United Kingdom Infrared Telescope. The approximate size and orientation of these outflows have been overlaid as magenta arrows on the $\beta$ map in Fig. \ref{fig:outflows}, with the location of the candidate driving source MHO 3240 (a YSO associated with the UKIDSS source GPS 43851 0698063) shown as a white star. By creating $\beta$ maps of outflows in L1157 using 1.3 mm MAMBO data and 850 \micron\ SCUBA data, \citet{Gueth+03} found that $\beta$ varied systematically across protostellar and outflow environments, suggesting that shocked regions may have smaller dust grains, or altered grain chemistry. They found that $\beta$ had its lowest value of $\sim 1.5$ at the location of the centre of the protostar in their observations, and reaching its highest value of $\sim 3.0$ in the shocked regions. In SDC24, we find a similar coincidence of this $\beta$ behaviour, albeit with an offset of $\Delta \beta \approx 0.5$. The candidate driving source MH 3240 falls on a minimum in our $\beta$ map, with a value of $\beta \approx 1.0$, and a maximum value of $\beta \approx 2.5$ is coincident with the tip of the \citet{Ionnidis+Froebrich12} southern outflow. However, we do not see any dust emission which is obviously associated with the outflow itself and we have insufficient evidence to definitively show that these $\beta$ variations are causally linked to the outflows from MHO 3240. Observations of shock tracers such as NH$_3$ or SiO could be acquired in the future for comparison with variations in the $\beta$ maps, and observations in, for example, CO (3$-$2) could identify any link between the molecular outflow and the IRDC.

In a recent study, adopting a similar techniques to this study with NIKA data, \citet{Bracco+17} found that out of three low-mass cores in the Taurus B213 filament, the two protostellar sources exhibited radially increasing $\beta$ profiles while the third pre-stellar core showed no evidence of any systematic variation in $\beta$. With neither SDC18 nor SDC24 showing any significant radial variations in our study, despite both being protostellar, how does this result fit into the combined picture with the \citet{Bracco+17} study? There are some important differences to consider because, although the angular scales measured are almost identical, SDC18 and SDC24 are far more massive and distant than the Taurus filament. At a distance of 140 pc, the 40\arcsec\ extent of the radial profiles corresponds to $\lesssim 0.03$ pc ($\lesssim 6000$ au), while it corresponds to $\gtrsim 0.8$ pc and $\gtrsim 0.6$ pc at the distances of SDC18 and SDC24, respectively. Assuming that $\beta$ increases radially with the slopes given in that study, we would not expect to be able to resolve such variation at the distances of SDC18 and SDC24. In addition, we expect the column densities in the two IRDCs to be much higher than in the Taurus filament, meaning that line-of-sight averaging is more significant in the former. \citet{Chacon-Tanarro+17} were also unable to resolve $\beta$ variations using NIKA in the pre-stellar Taurus core L1544, even though their resolution matches that of \citet{Bracco+17}.

\subsection{Mass concentration}

SDC18 and SDC24 are morphologically very different when observed at NIKA resolution, which can be seen on the high-angular resolution column density maps of Fig. \ref{fig:columndensity}. One striking feature is the large number of compact sources of SDC18 revealed in the NIKA images. By contrast, the column density structure in SDC24 is relatively simple, with a dominant central clump surrounded by a plateau of more diffuse emission. The 8 \micron\ GLIMPSE image (see Fig. \ref{fig:rawdata}) shows that there is a more complex underlying filamentary structure below the spatial sensitivity of the dust continuum data, but which must contain a small amount of the mass when compared to the central clump. The morphology of these two IRDCs also differ by their aspect ratios -- cf. Fig. \ref{fig:massconcentrationB} b) --  where SDC18 is more elongated and filamentary than SDC24. 

Fragmentation theories of gravitationally unstable clouds have shown that there are key differences in the wavelength of the fastest growing mode of density perturbations depending on the initial morphology of the cloud \citep{Larson85}. While for spherical clouds, density perturbations of the length scale of the cloud diameter grow the fastest, while for filaments, perturbations of only a fraction of the filament's length develop more rapidly \citep{Inutsuka+Miyama92,Clarke+16}. As a consequence, filamentary clouds will more easily fragment into multiple cores than more spherical clouds that will concentrate most of their mass into a single fragment, naturally leading to a larger mass concentration. The differences we observe between SDC18 and SDC24 could therefore be explained by different initial conditions regarding the morphology of their respective mass reservoirs.

\begin{figure}[t]
	\centering
	\includegraphics[width=\hsize]{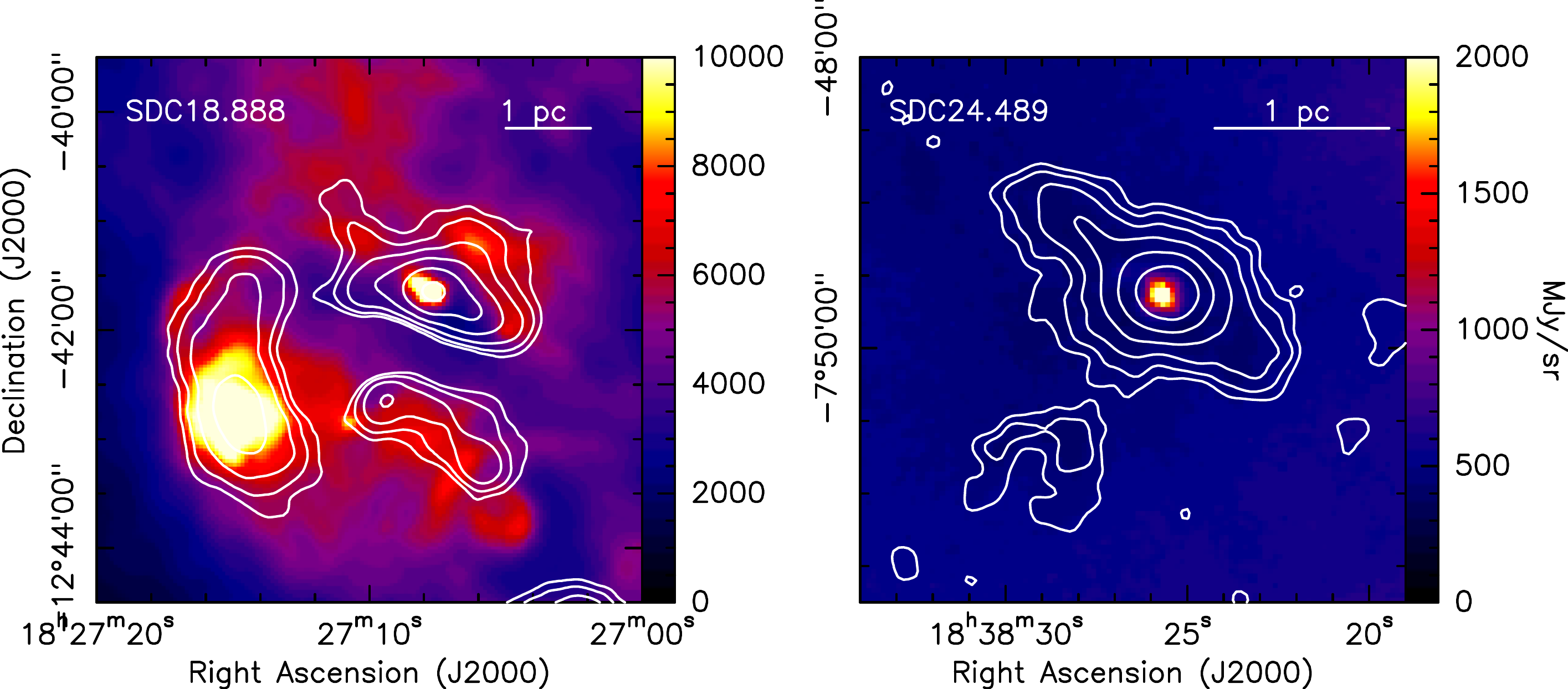}
    \caption{\textit{Herschel} 70 \micron\ images of SDC18 (left) and SDC24 (right) from the Hi-GAL survey. Note the bright 70 \micron\ sources at the locations of the column density peaks. The contours show the 2.0 mm S/N as in Fig. \ref{fig:ratiomaps}.}    	\label{fig:70micron}
\end{figure}

We could also envisage that the observed differences are due to different cloud evolutionary stages. We could imagine that SDC24 started off more filamentary (as SDC18) and that as a result of gravitational contraction ended up more spherical. If this would be correct, then SDC24 should have been forming stars for many free-fall times and should host a rich cluster at its centre, as predicted by the recent models of \citet{Vazquez-Semadeni+17}. At 8 \micron\ there is no evidence of such cluster, but we do notice that the column density peaks are well matched with a compact source seen at 70 \micron\ in Fig. \ref{fig:70micron} (sources HIGALPB018.8888-0.4744 and HIGALPB024.4888-0.6914 in the \citet{Molinari+16} catalogue) and 70 \micron\ fluxes also appear to be in excess compared to the SEDs of the cold clumps in Fig. \ref{fig:SED}. Luminosity at 70 \micron\ has been shown to correlate well with internal protostellar luminosity, $L_\mathrm{int}$, \citep{Dunham+08} according to:

\begin{equation}
	L_\mathrm{int} = 3.3 \times 10^8 \, F_{70}^{0.94} \, \left(\frac{d}{140 \, \mathrm{pc}}\right)^2 L_\odot,
\end{equation}  

\noindent where the 70 \micron\ flux, $F_{70}$, has units of ergs cm$^{-2}$ s$^{-1}$, and $d$ is the distance to the source in parsecs. Using the 70 \micron\ fluxes from the \citet{Molinari+16} catalogue, we estimated the luminosity of young stellar objects seen at 70 \micron\ for both IRDCs, finding $L_\mathrm{int} = 1410 \pm 170$ \lsol\ for SDC18 and $155 \pm 18$ \lsol\ for SDC24. Therefore, at the zero order, SDC18 is forming a larger number and/or more massive protostars than SDC24, validating the evolution scenario. However, because SDC18 is more massive, we would expect it to form more stars anyway. Once the luminosities are normalised by the cloud masses, we obtain the following luminosity to mass ratios: $L_\mathrm{int}/M = 0.22 \pm 0.04 \, \mathlsol \, \mathmsol^{_-1}$ for SDC18 and $L_\mathrm{int}/M = 0.25 \pm 0.06 \, \mathlsol \, \mathmsol^{_-1}$ for SDC24. The similar L/M values suggest that in absolute terms, both clouds are forming a comparable number of stars per unit mass of gas (assuming the same star formation efficiency applies in both regions), and would therefore seem to exclude the possibility of different evolutionary stages as the cause of the morphological differences between SDC18 and SDC24. 

\subsection{Future studies}

We have shown that creating maps of the dust emissivity spectral index, $\beta$, can be achieved with NIKA data with relatively high resolution and precision. The absolute calibration of NIKA data, however, presents the biggest challenge in terms of determining the absolute value of $\beta$, and improvement on this front could allow variation from region-to-region to be studied.

NIKA was a pathfinder instrument, in service on the IRAM 30-m telescope from 2010--2015, and its much larger successor, NIKA2 \citep{Adam+18}, has been recently commissioned. Further studies into $\beta$ variations will be carried out as part of the GASTON survey (Galactic Star Formation with NIKA2), a guaranteed-time large programme with NIKA2, and improvements to the absolute calibration can be made by using the low-resolution spaced-based imaging from \textit{Planck} to define the background levels. Part of this programme will include a high-sensitivity study of part of the Inner Galactic plane in a region rich with high-mass star-forming clumps and giant molecular filaments, allowing studies of both $\beta$ and mass concentration, as in this paper, to be expanded to a much larger and uniformly observed sample.

\section{Summary} \label{sec:Summary}

We have used 1.2 mm and 2.0 mm imaging from NIKA on the 30 m IRAM telescope to study two IRDCs from the catalogue of \citet{Peretto+Fuller09}, SDC18 and SDC24. While SDC24 lies in a quiescent region, we find that some of the 2.0 mm emission in the SDC18 region may be heavily contaminated with free--free emission from the W39 \ion{H}{II} region, and we use the NIKA imaging in tandem with data from the Hi-GAL survey to determine dust properties surrounding the most massive clumps in both regions.

We have constructed maps of the dust emissivity spectral index, $\beta$, from the ratio of the specific intensities in both NIKA2 wavebands after convolving them to a common angular resolution. We use a FFT-based PSF-matching procedure to match the 1.2 mm NIKA PSF to the 2.0 mm PSF, finding that the telescope beam sidelobes, if not accounted for, can introduce artificial structures into the $\beta$ maps, particularly in the neighbourhoods of bright compact sources. The weighted mean central values for $\beta$ in SDC18 and SDC24 are $\bar{\beta} = 2.07 \pm 0.09$ (random) $\pm 0.25$ (systematic) and $\bar{\beta} = 1.71 \pm 0.09$ (random) $\pm 0.25$ (systematic), respectively. Although we find no evidence for significant systematic radial variations of $\beta$ in the dust immediately surrounding the dense clumps in the centres of these two IRDCs, we note that sufficiently high S/N are achieved over very limited areas, with radii $\lesssim 0.3$ pc. Deviations from the mean $\beta$ value in SDC24 might be linked to molecular hydrogen outflows from a nearby YSO, though we do not have sufficient evidence to make a definite causal link. In addition, since the observations of both sources were carried out on the same night, the same systematic calibration offsets apply and the relative difference in the central $\beta$ values are reliable and significant. We suggest that the elevated $\beta$ value in SDC18 compared to SDC24 may be caused by its proximity to the W39 \ion{H}{ii} region.

We have also studied the relative mass concentration within the two IRDCs using a dendrogram-based analysis on high-resolution (13\arcsec) maps of the H$_2$ column density. We find that SDC24 is more centrally concentrated than SDC18, with a higher fraction of its mass concentrated within smaller normalised radii and lower aspect ratios at all column density contour levels. Both IRDCs have the same protostellar luminosity per unit of mass, indicating that the morphological differences can not simply  be attributed to an age effect, whereby SDC24 could have been undergoing gravitational collapse for longer than SDC18. This evidence supports a more spherical initial state for SDC24 than SDC18, with internal fragmentation occurring on longer timescales than global collapse, while the more filamentary SDC18 may be more unstable to fragmentation on smaller scales.

Future studies of high-mass IRDCs using the techniques described in this paper with NIKA's successor, NIKA2, as part of the GASTON large programme, will allow relative variations in $\beta$ to be measured over larger areas. In addition, with the higher sensitivities that will be achieved, we will be able to study the mass concentration over a much larger sample of IRDCs, allowing more quantitative conclusions about their initial conditions to be reached.

\begin{acknowledgements}

We thank the anonymous referee for helpful comments which have improved the quality and clarity of this paper. A. J. R. would like to thank the STFC for postdoctoral support under the consolidated grant number ST/N000706/1. N. P. also wishes to acknowledge support under the above STFC consolidated grant as well as further support from the STFC under grant number ST/M000893/1. We thank the Royal Society for providing computing resources under Research Grant number RG150741.  We would like to thank the IRAM staff for their support during the campaigns. The NIKA dilution cryostat has been designed and built at the Institut N\'eel. In particular, we acknowledge the crucial contribution of the Cryogenics Group, and in particular Gregory Garde, Henri Rodenas, Jean Paul Leggeri, Philippe Camus. This work has been partially funded by the Foundation Nanoscience Grenoble, the LabEx FOCUS ANR-11-LABX-0013 and the ANR under the contracts "MKIDS", "NIKA" and ANR-15-CE31-0017. This work has benefited from the support of the European Research Council Advanced Grant ORISTARS under the European Union's Seventh Framework Programme (Grant Agreement no. 291294). We acknowledge fundings from the ENIGMASS French LabEx (R. A. and F. R.), the CNES post-doctoral fellowship program (R. A.),  the CNES doctoral fellowship program (A. R.) and the FOCUS French LabEx doctoral fellowship program (A. R.). This research made use of astrodendro, a Python package to compute dendrograms of Astronomical data (http://www.dendrograms.org/), as well as \textit{Astropy}, a community-developed core Python package for astronomy \citep{astropy}, \textit{SciPy} \citep{scipy} and \textit{matplotlib} \citep{Matplotlib}. This research has also made use of the NASA Astrophysics Data System. 

\end{acknowledgements}

\bibliographystyle{aa}
\bibliography{References_NIKA_IRDC}

\begin{appendix}

\section{Sources of contamination} \label{sec:contamination}
\subsection{Free-free contamination} \label{sec:freefreecontamination}

We modeled the 20 cm continuum emission as an optically thin \ion{H}{II} region, with an SED described by the power law relation $S_\ion{H}{II}(\nu)\propto \nu^{-0.1}$ \citep[e.g.][]{Mezger+Henderson67,Wilson+12}, in order to estimate the level of contamination from the 20 cm continuum emission in both NIKA bands. After smoothing the data to 20 arcsec resolution, the flux density was calculated for both NIKA bands at 1.2 mm and 2.0 mm. Maps of the ratios of the modelled \ion{H}{II} region flux density at 1.2 mm and 2.0 mm to the flux density of the corresponding NIKA maps are shown in Fig. \ref{fig:contamination}. It can be seen that, while the brightest source to the north-west is contaminated at less than the 20\% level in both bands at its centre, the source to the east is dominated by free--free emission, especially at 2.0 mm -- so much so that it may not be a dust continuum source at all. The contamination fraction exceeds unity in many areas in the 2.0 mm image, indicating that we have overestimated the 2.0 mm flux from the free--free emission and that the assumed SED is not describing these observations particularly well. We did not, therefore, subtract the calculated contaminant flux from our NIKA images, but we regard region C as free--free dominated for the purpose of this study.

\begin{figure}
	\centering
	\includegraphics[width=\hsize]{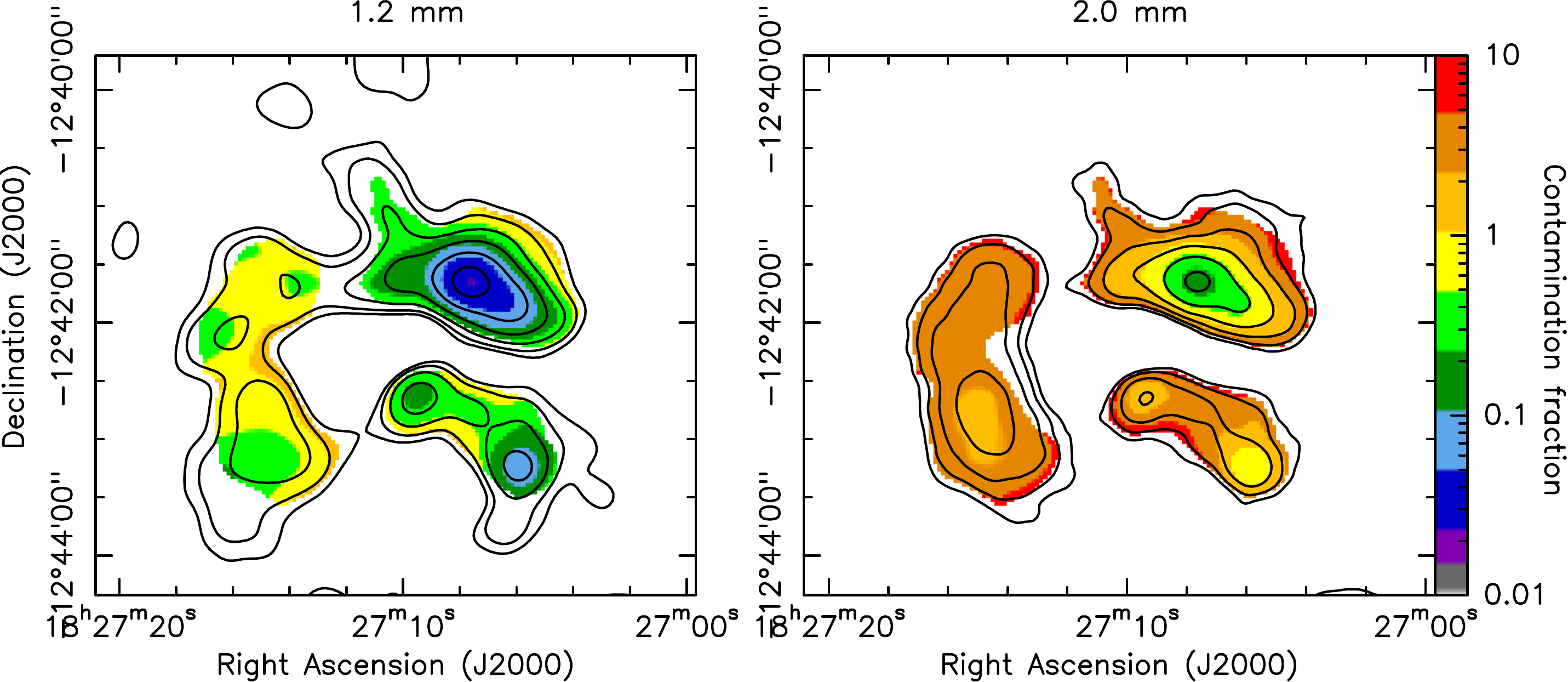}
	\caption{Maps of the free--free contamination fraction at 1.2 mm and 2.0 mm, derived from the 20 cm MAGPIS data assuming an optically thin \ion{H}{II} region. The black contours overlaid are from the corresponding NIKA maps at 20 arcsec resolution, and identical to those displayed in Fig.~\ref{fig:ratiomaps}.}
	\label{fig:contamination}
\end{figure}

To fully account for the contamination, a better characterisation of the contaminant \ion{H}{II} region is necessary and, while the MAGPIS data have excellent angular resolution, we lack data at other radio wavelengths to properly constrain the SED of the \ion{H}{II} region. For future studies, data from the \ion{H}{I}/OH/Recombination line survey of the inner Milky Way \citep[THOR;][]{Beuther+16} could be used to determine the nature of the 20 cm continuum emission (i.e. free-free vs. synchrotron emission), allowing the contamination to be more accurately modelled at 20 arcsec resolution and subtracted from the 1.2 mm and 2.0 mm imaging. Additional continuum imaging in, for example, the 6 cm band would ensure that \ion{H}{ii} regions could also be caught in the optically thin regimes, which will not necessarily be true at 20 cm.

\subsection{Line contamination} \label{sec:COcontamination}

There is also potential for contamination in our 1.2 mm NIKA images from the $^{12}$CO (2$-$1) emission line at 230.538 GHz, which lies towards the low-frequency edge of the 1.2 mm band \citep[see Figure 2. of][]{Bracco+17}. In their NIKA study, \citet{Bracco+17} found this contamination to be at the level of $\sim 1\%$ towards their target filament in Taurus by using archival data from the Five College Radio Astronomical Observatory (FCRAO), and we expect a higher level of contamination given that our targets lie on the Inner Galactic plane. Since no archival CO (2$-$1) data exists towards our targets, we estimated the level of contamination in our data by using $^{13}$CO (1$-$0) data from the Boston University FCRAO Galactic Ring Survey \citep[GRS][]{Jackson+06}. The GRS data have an angular resolution of 46\arcsec\ and so we smoothed both our 1.2 mm imaging and the GRS data to an effective angular resolution of 50\arcsec, and integrated the GRS cubes over the full velocity range of $-5$ to 135 km s$^{-1}$ to generate maps of integrated intensity in units of K km s$^{-1}$, and reprojected them onto a pixel grid matching our observations. 

Converting an integrated intensity map of $^{13}$CO (1$-$0) emission into $^{12}$CO (2$-$1) requires a number of assumptions to be made. In conditions of local thermodynamic equilibrium, a common excitation temperature is shared between all CO isotopologues, and the ratio of brightness temperatures of the same rotational transition of $^{12}$CO and $^{13}$CO are related by:

\begin{equation}
	\frac{T_\mathrm{b}(^{12}\mathrm{CO})}{T_\mathrm{b}(^{13}\mathrm{CO})} = \frac{1 - e^{-\tau(^{12}\mathrm{CO})}}{1 - e^{-\tau(^{13}\mathrm{CO})}},
\end{equation}

\noindent where the optical depths, $\tau$, for each isotopologue are related by their abundance ratio:

\begin{equation}
	\tau(^{12}\mathrm{CO}) = a_{12/13} \, \tau(^{13}\mathrm{CO}),
\end{equation}

\noindent where $a_{12/13}$ is the ratio of the abundance of $^{12}$CO compared to $^{13}$CO. Isotopic abundances are measured to change as a function of Galactocentric distance, and we adopted abundance ratios of $a_{12/13} = 40$ for SDC18 and $a_{12/13} = 45$ for SDC24, following the prescription of \citet{Wilson+Rood94}. We identified the molecular cloud counterparts to SDC18 and SDC24 in the catalogue of \citet{Roman-Duval+10} as G018.89-00.51 and G024.49-00.71, respectively, and by adopting the mean $^{13}$CO optical depths they derived for each cloud, we converted the $^{13}$CO (1$-$0) integrated intensity maps to maps of integrated $^{12}$CO (1$-$0) intensity. We converted the maps of $^{12}$CO (1$-$0) into maps of $^{12}$CO (2$-$1) by multiplying by a line ratio of $R_{2-1/1-0} = 0.7$, which is appropriate for cold, dense molecular gas \citep{Penaloza+17}.

The GRS data contain no inherent filtering of spatial frequencies, since sky subtraction for spectral data is carried out by using emission-free frequency channels. We therefore applied the 1.2 mm NIKA transfer function (see Sect. \ref{sec:SED} for a description of the transfer function) to the GRS data to apply spatial filtering that is equivalent to that applied to the NIKA continuum data, thus allowing the images to be compared directly. Finally, we followed the procedure of \citet{Drabek+12} to convert the maps of $^{12}$CO (1$-$0) integrated intensity into units of \mjybeam, by integrating over the 1.2 mm NIKA bandpass, and thereby into \mjysr\ for direct comparison to our 1.2 mm images.

We found that the contamination is generally $\sim 1-3\%$, with the smaller values occurring with the peaks of 1.2 mm emission. CO (2$-$1) is likely to be optically thick for a Galactic plane cloud due to its relatively low critical density and so this contamination will be less important at areas of high column density. However, this contamination could become stronger at lower column densities, introducing a systematic bias as a function of column density into the 1.2 mm to 2.0 mm intensity ratio as an extreme case. Although the angular resolution is necessarily reduced to 50\arcsec\ for this test, we have seen no evidence that CO contamination is significant in our targets.

\section{Synthetic $\beta$ maps} \label{sec:syntheticbetamaps}

To determine what kind of artificial morphology might be introduced into the $\beta$ maps as a consequence of noise and PSF effects we created synthetic $I_1/I_2$ ratio maps for two source models: i) a perfect point-source and ii) a Gaussian source with a FWHM of 19\arcsec, approximately equal to the deconvolved FWHM of the main clump in SDC24. To create synthetic 1.2 mm and 2.0 mm emission maps, each model was created on a 2\arcsec\ pixel grid and convolved with a beam map (measured on Uranus and described in Sect. \ref{sec:crkernel}) before Gaussian noise was added at the S/N level measured for SDC24, listed in Table \ref{tab:noise}. These synthetic raw images were then smoothed to an effective resolution of 20 arcseconds using the method described in Sect \ref{sec:crkernel} before the 1.2 mm and 2.0 mm images were rescaled such that the ratio at the intensity peak corresponds to the ratio corresponding to $\beta = 1.8$ and $T_\mathrm{d} = 14.0$ K according to Eq. \ref{eq:ratio}. To explore the effects of beam asymmetries, a second synthetic observation was also produced for both source models using the same noise realisations at 1.2 mm and 2.0 mm, but differing in that the beam map was azimuthally averaged before the beam convolution. After masking to include only pixels with a S/N > 3 in both wavebands, the synthetic observations were used to produce $\beta$ maps assuming a dust temperature of $T_\mathrm{d} = 14.0$ K, matching the input, so that the $\beta$ pixel values could be compared directly to the input value.

\begin{figure}
	\centering
	\includegraphics[width=\hsize]{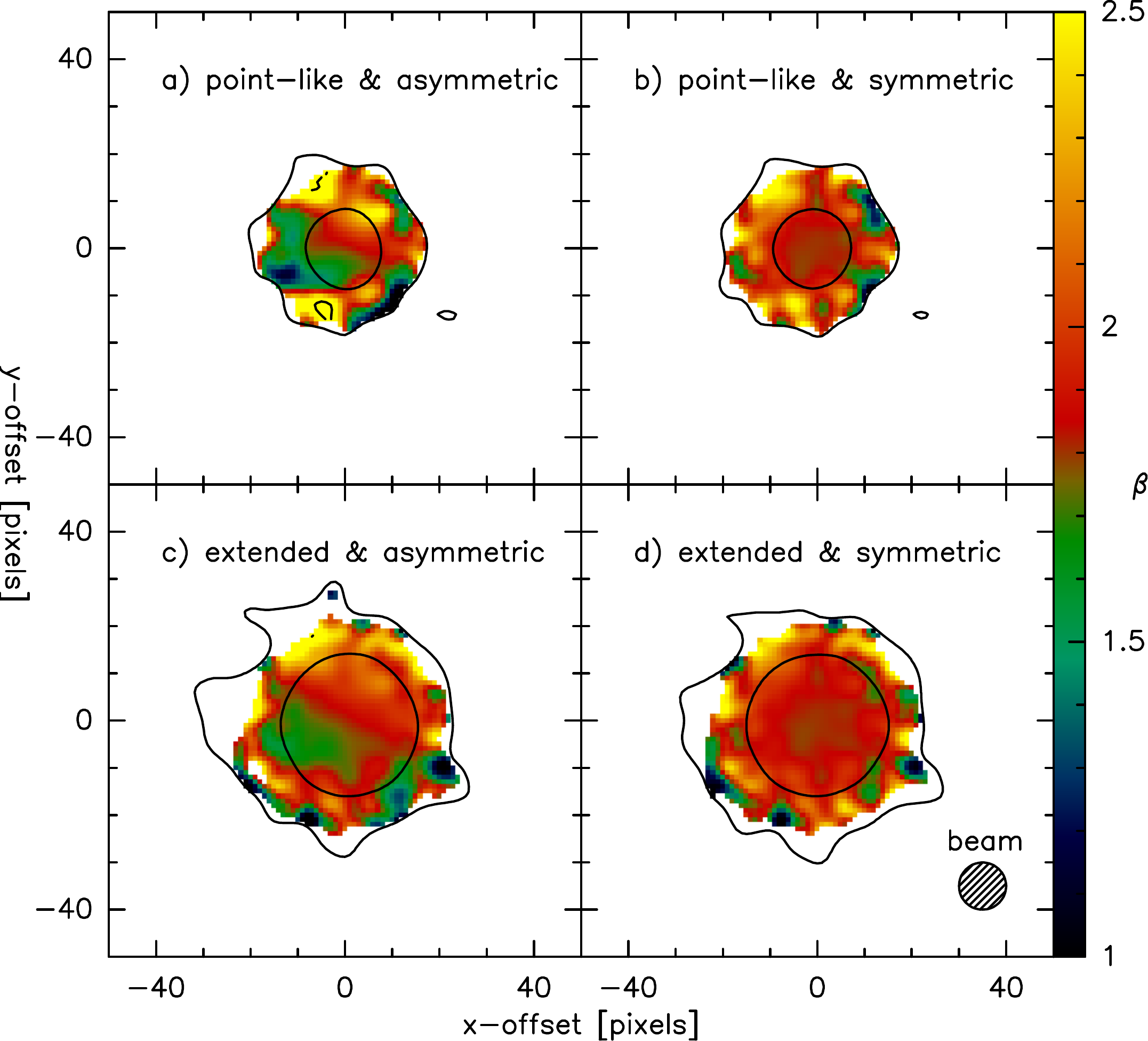}
	\caption{Synthetic $\beta$ maps generated for two separate models: a) and b) are generated for a point source and c) and d) are generated for a source with a Gaussian intensity profile with a FWHM of 19 arcseconds. Panels a) and c) have been generated by convolution of the source with a Uranus beam map, while panels b) and d) were convolved with the beam map after azimuthal averaging. The effective beam size is given in the lower-right of panel d), and the contours represent S/N levels of 3 and 25 from the 1.2 mm image.}
         \label{fig:betamodels}
\end{figure}

The four synthetic $\beta$ maps are illustrated in Figure \ref{fig:betamodels}, with versions created using the observed (asymmetric) beam map as well as the azimuthally-averaged (symmetric) beam map for each of the two source models. In all cases, the presence of noise has introduced a host of synthetic features that deviate far from the input $\beta$ value, with the most significant deviations at the edges of the maps. This would be expected as any coincidence as any extrema in the noise in either or both wavebands could translate into very large spurious deviations in their ratio after smoothing. To further constrain which regions of the $\beta$ maps can be trusted, an uncertainty map for the noise contribution to $\beta$ can be generated from the two S/N maps:

\begin{equation} \label{eq:betaerror}
	\Delta \beta_\mathrm{noise} = 1.82 \sqrt{\frac{1}{(S/N)_1^2} + \frac{1}{(S/N)_2^2}}
\end{equation}

\noindent where $(S/N)_1$ and $(S/N)_2$ are the signal-to-noise ratios in the 1.2 mm and 2.0 mm bands, respectively, and the numerical factor is $[\ln\,(\nu_1/\nu_2)]^{-1}$. The accuracy to which $\beta$ can be determined, based upon noise arguments alone and provided that a certain S/N has been achieved in both wavebands, can be described by $\Delta \beta_\mathrm{noise} = 2.57/\mathrm{(S/N)}$; to restrict noise-based uncertainties to $\Delta \beta_\mathrm{noise} = 0.1$, we must look only at pixels for which the S/N is $\gtrsim 25$ in both wavebands. The black contours in Fig. \ref{fig:betamodels} correspond to S/N of 3 and 25 for the synthetic 1.2 mm emission, and so we should expect to see noise-based deviations of $\Delta \beta_\mathrm{noise} < 0.1$ within the top contour. This is approximately true in the case of the models that used a circularly-symmetric beam pattern, but more significant deviations can be seen in the more realistic asymmetric case, indicating that  asymmetries in the beam that may arise as a result of pointing or focus errors can induce significant artificial structures.

Comparison of the artificial $\beta$ maps of Fig. \ref{fig:betamodels} with the measured $\beta$ maps of Fig. \ref{fig:betamap} then allows the measured structures to be placed in some context. The white contour in Fig. \ref{fig:betamap} marks the boundary above which the 1.2 mm S/N is greater than 25, where we can therefore expect to start detecting any significant $\Delta \beta$ structures. The areas in the mapping where this criterion is met are very limited, spanning only a slightly larger area than one beam width and any underlying structure is certainly unresolved. This comparison indicates that care should therefore be taken when interpreting features that fall below such a contour of S/N $\sim 25$, though we should be confident in any variations detected above this level.
\end{appendix}

\end{document}